\documentclass[11pt,a4paper]{article}
\usepackage[T1]{fontenc}
\usepackage[utf8]{inputenc}
\usepackage{authblk}
\usepackage{fullpage}
\usepackage{graphicx}
\usepackage{xcolor}
\usepackage{algorithm}
\usepackage{algpseudocode}
\usepackage{amssymb}
\usepackage{amsthm}
\usepackage{amsmath}
\usepackage{url}
\usepackage{bbold}
\usepackage{stmaryrd}
\usepackage{txfonts}
\usepackage[square,numbers,comma,sort&compress]{natbib} 
\usepackage[colorlinks,urlcolor=red,citecolor=red,linkcolor=blue]{hyperref}
\newtheorem{thm}{Theorem}
\newtheorem{lem}[thm]{Lemma}
\newtheorem{cor}[thm]{Corollary}
\newtheorem{defi}{Definition}
\newtheorem*{rmk}{Remark}
\newtheorem*{pf}{Proof}
\title{FFT-based Kronecker product approximation to micromagnetic long-range interactions}
\author[1]{Lukas Exl\thanks{Corresponding author, \texttt{lukas@exl.at}, \texttt{lukas.exl@fhstp.ac.at}}}
\author[2,3]{Claas Abert}
\author[4]{Norbert J. Mauser}
\author[1]{Thomas Schrefl}
\author[4]{Hans Peter Stimming}
\author[5]{Dieter Suess}

\affil[1]{University of Applied Sciences, Department of Technology, A-3100 St.Poelten, Austria}
\affil[2]{Fachbereich Mathematik, Universit\"at Hamburg, Bundesstr. 55, 20146, Hamburg, Germany}
\affil[3]{Institut f\"ur Angewandte Physik und Zentrum f\"ur Mikrostrukturforschung, Universit\"at Hamburg, Jungiusstr. 11, 20355 Hamburg, Germany}
\affil[4]{Wolfgang Pauli Institute c/o Fakult\"at f. Mathematik, Universit\"at Wien, A-1090 Vienna, Austria}
\affil[5]{Institute of Solid State Physics, Vienna University of Technology, A-1040 Vienna, Austria}

\begin{document}
\maketitle
\begin{abstract}
We derive a Kronecker product approximation for the micromagnetic long range interactions in a collocation framework by means of separable sinc quadrature. Evaluation of this operator for structured tensors (Canonical format, Tucker format, Tensor Trains) scales below linear in the volume size. Based on efficient usage of FFT for structured tensors, we are able to accelerate computations to quasi linear complexity in the number of collocation points used in one dimension. Quadratic convergence of the underlying collocation scheme as well as exponential convergence in the separation rank of the approximations is proved. Numerical experiments on accuracy and complexity confirm the theoretical results.
\end{abstract}
\section{Introduction}
Micromagnetism is a continuum theory for the treatment of magnetization processes in ferromagnetic bodies \cite{brown_1963}. In the investigation of ferrormagnetic materials micromagntetic simulations are nowadays of high interest \cite{fidler_2000} and play an essential role in the development of magnetic data storage devices \cite{schabes_micromagnetic_2008}. Starting from the total Gibb's free energy functional, containing exchange, magnetocrystalline anisotropy and magnetostatic contributions, micromagnetics either solves for local minimum configurations or solves an equation of motion for the magnetization \cite{brown_1963}. Among these components the magnetostatic field is the only non-local term and gives rise to magnetization structures on a length scale which is orders of magnitude greater than atomic spacing. 
Naive implementation of the superposition-based integral operators \eqref{intop2} or solvers for the underlying differential equation (Poisson equation \eqref{poisson}) yield computational costs proportional to the square of the number of grid points, i.e. $\mathcal{O}(N^2)$. Since the spacial resolution in such numerical computations has to be high enough for the correct description of magnetic domains, a quadratic scaling is almost never feasible. Historically, different methods have been proposed to reduce the computation effort. Juan and Bertram \cite{yuan_fast_1992} used the fast Fourier transform for evaluating the convolution of the magnetization with demagnetizing tensor. Blue and Scheinfein \cite{blue_1991} applied a multipole expansion to the integration kernel. Both methods reduce the computational effort for magnetostatic field evaluation to $\mathcal{O}(N\log N)$. Fredkin and Koehler \cite{fredkin_1990} coupled the finite element method with a boundary integral method to treat the open boundary problem in micromagnetics, cf. \eqref{poisson}. If hierarchical matrices are used to compress the boundary element matrix \cite{forster_fast_2003}, the computation of the boundary element part will be $\mathcal{O}(m\log m)$, where $m$ is the number of nodes at the surface.\\
Based on developments of approaches addressing high dimensional problems \cite{beylkin_2002},\cite{ hackbusch_low-rank_2005} with a solution that can be approximated by separable functions, recently also tensor approximation methods \cite{goncharov_2010},\cite{exl_fast_2012_2} were introduced into micromagnetics.\\ 
In this paper we give a detailed mathematical analysis of the \textit{tensor grid method} introduced in \cite{exl_fast_2012_2} and extend it to a FFT-based method.\\[0.1cm]
In micromagnetics, the magnetization $\boldsymbol{m}$ (a vector field in a closed and bounded domain $\Omega \subset \mathbb{R}^3$, which is zero outside) generates the so-called \textit{stray field} or \textit{demagnetizing field} $\boldsymbol{h}_d = -\nabla \phi$. The \textit{scalar potential} $\phi$ is the solution to \cite{jackson_classical_1999}
\begin{align}\label{poisson}
  \Delta \phi = \nabla \cdot \boldsymbol{m},
\end{align}
whereby $|\phi| = \mathcal{O}(1/r)$ and $|\nabla \phi| = \mathcal{O}(1/r^2)$ as the distance $r\rightarrow \infty$, \cite{brown_1963}. These regularity conditions are often referred to as open boundary conditions \cite{fidler_2000}.\\
From the fundamental solution to the Laplace operator in $\mathbb{R}^3$ it is possible to express $\phi$ as integral representation, i.e.
\begin{align}\label{intop1}
  \phi(\boldsymbol{x}) =
  - \frac{1}{4\pi} \int_\Omega \frac{\nabla \cdot \boldsymbol{m}(\boldsymbol{y})}{\left\|\boldsymbol{x} - \boldsymbol{y}\right\|} \text{d} \boldsymbol{y}
  + \frac{1}{4\pi} \int_{\partial \Omega} \frac{\boldsymbol{n}(\boldsymbol{y}) \cdot \boldsymbol{m}(\boldsymbol{y})}{\left\|\boldsymbol{x} - \boldsymbol{y}\right\|} \text{d} \sigma^\prime.
\end{align}
Integration by parts yields 
\begin{align}\label{intop2}
\phi(\boldsymbol{x}) =
  - \frac{1}{4\pi} \int_\Omega \boldsymbol{m}(\boldsymbol{y}) \cdot \frac{\boldsymbol{x} - \boldsymbol{y}}{\left\|\boldsymbol{x} - \boldsymbol{y}\right\|^3} \text{d} \boldsymbol{y}.
\end{align} The latter expression makes sense in micromagnetics due to the constraint $|\boldsymbol{m}| = 1$ a.e. in $\Omega$ \cite{brown_1963}, which implies $\boldsymbol{m}\in L^\infty (\Omega)$. Since the kernels $\kappa^{(q)}(\boldsymbol{x}) := x^{(q)}/\left\|\boldsymbol{x}\right\|^3 \in L^p\big(B_{R}(\boldsymbol{0})\big)$ for balls $B_{R}(\boldsymbol{0})$ centered in $\boldsymbol{0}$ with $R>0$ and $p \in [1,3/2)$, H\"older's inequality ensures that \eqref{intop2} is well-defined.\\[0.1cm]
 The paper is organized as follows. In Sec.~\ref{formats} we give an introduction into the two widely used tensor formats, i.e. \textit{canonical tensors} and \textit{Tucker tensors}, where we also focus on aspects like (best) approximation from a theoretical and practical point of view. In Sec.~\ref{sec_kron} we prove quadratic convergence of a collocation scheme for the micromagnetic potential operator and derive a Kronecker product approximation, which is proved to be exponentially convergent. A brief description on \textit{sinc-function based approximation theory} is also given, as well as aspects of recompression of the separable approximation. Sec.~\ref{fft} deals with the FFT acceleration of the potential evaluation, where we derive quasi linear complexity. Numerics in the end of the section confirm these results. 
\section{Tensor formats}\label{formats}
For an extensive review on structured tensors and some algorithms to compute structured tensor approximations see \cite{kolda_tensor_2009} and references therein. The most common arithmetic operations on Tucker and canonical tensors are presented in \cite{bader_efficient_2008}; in addition we give a description of the \textit{Hadamard product} and TT tensors \cite{oseledets_2011}, Sec.~\ref{arithmetics}. \\
The following explanations are carried out for order-$3$ tensors, since the generalization to higher order tensors is straight forward.\\
We denote the set of (order-$3$) tensors
\footnote{Another common notation is $\mathbb{K}^{\boldsymbol{I}}$ where $\boldsymbol{I}=I_1 \times I_2 \times I_3$ and $I_p = \{1 \hdots n_p\}$, see \cite{kolda_tensor_2009}}
with mode sizes $\boldsymbol{n} = (n_1,n_2,n_3)$ over the field $K = \mathbb{R}\,\,\text{or}\,\,\mathbb{C}$ with $\bigotimes_{p=1}^3\mathbb{K}^{n_p}$ and the set of matrices of size $n_1 \times n_2$ as usual with $\mathbb{K}^{n_1 \times n_2}$.\\ 
Let $\mathcal{X},\mathcal{Y} \in \bigotimes_{p=1}^3\mathbb{R}^{n_p}$. The \textit{Frobenius norm} is defined as 
\begin{align}
\left\| \mathcal{X} \right\|_F := \sqrt{\sum_{i_1=1}^{n_1} \sum_{i_2=1}^{n_2} \sum_{i_3=1}^{n_3} x^2_{i_1 i_2 i_3}},
\end{align}
which is associated with a scalar product 
\begin{align}
\left\langle \mathcal{X},\mathcal{Y} \right\rangle := \sum_{i_1=1}^{n_1} \sum_{i_2=1}^{n_2} \sum_{i_3=1}^{n_3} x_{i_1 i_2 i_3}\,y_{i_1 i_2 i_3},
\end{align}
with $\left\| \mathcal{X} \right\|_F^2 = \left\langle \mathcal{X},\mathcal{X} \right\rangle$.
\subsection{Canonical tensors}\label{Canonical}
A tensor $\mathcal{X} \in \bigotimes_{p=1}^3\mathbb{R}^{n_p}$ is said to be in
\textit{canonical format (CANDECOMP/PARAFAC (CP) decomposition)} with \textit{(outer product) rank}\, $ R $, if
\begin{align} \label{CP}
\mathcal{X} = \sum_{r=1}^{R} \lambda_r\,\,u^{(1)}_r \circ u^{(2)}_r \circ u^{(3)}_r,
\end{align}
with $\lambda_r \in \mathbb{R} $,\, (unit) vectors $u^{(j)}_r \in \mathbb{R}^{n_j}$,
and $\circ$ is the tensor outer product.
Abbreviating notation as in \cite{kolda_tensor_2009}, a tensor in CP format is written as
\begin{align} \label{CP2}
\mathcal{X} \equiv \llbracket \boldsymbol{\lambda};\,U^{(1)},U^{(2)},U^{(3)} \rrbracket,
\end{align}
with weight vector $\boldsymbol{\lambda} = [ \lambda_1,\ldots,\lambda_R ] \in \mathbb{R}^{R}$
and (factor) matrices $U^{(j)} = \big[\,u^{(j)}_1 \,|\, \hdots \,|\, u^{(j)}_R\,\big] \in \mathbb{R}^{n_j \times R}$.
The storage requirement for the canonical tensor format amounts to $R\,\sum_{j=1}^{3} n_j$.\\
In the following we write $\mathcal{C}_{\textbf{n},r}$ for the set of canonical tensors with mode sizes $\textbf{n} = (n_1, n_2, n_3)$ and rank $r$, and simple $\mathcal{C}_{n,r}$, when the mode sizes are equal.\\
The \textit{tensor (outer-) product rank}, i.e. the minimal number of rank-$1$ terms in a representation like \eqref{CP} for a tensor $\mathcal{X}$, is an analogue of the matrix rank, however there are major differences between those two \cite{kolda_tensor_2009}. The product rank of a tensor might be different over $\mathbb{R}$ and $\mathbb{C}$; in principle there is no easy algorithm to determine the tensor rank since this is an NP-complete problem \cite{Haastad_1990}. In fact, there are several specific examples of tensors where only bounds exist for their ranks.\\
Moreover the tensor rank is not \textit{upper semicontinuous}, e.g. there exist sequences of tensors of rank $\leq r$ converging to a tensor of rank greater than $r$ \cite{de_silva_tensor_2008}. There is no \textit{Eckart-Young Theorem} available, i.e. a CP decomposition can not be computed via the SVD, indeed, it is possible that the best rank-$r$ approximation of a tensor of order greater than two may not even exist for the case $r>1$, \cite{de_silva_tensor_2008}.\\
Nevertheless a broad community uses canonical decomposition, e.g. \textit{psychometrics}, \textit{data mining}, \textit{neuroscience}, \textit{image compression and classification}, see \cite{kolda_tensor_2009} references theirin.\\
Algorithms for computing canonical decompositions are mostly based on optimization, e.g. \textit{alternating least squares} \cite{kolda_tensor_2009}, \textit{gradient based} or \textit{nonlinear least squares methods} \cite{acar_scalable_2011} or \textit{Gauss-Newton} \cite{tomasi_2006}.\\ 
Also approximation of \textit{operators} like the \textit{multiparticle Schr\"odinger operator} \cite{beylkin_2002} or Newton potential \cite{hackbusch_low-rank_2005} can be done by using the canonical tensor format in order to overcome the \textit{curse of dimensionality}.\\
For matrices $A \in \mathbb{R}^{\big(\prod_{i=1}^3 n_i\big) \times \big(\prod_{i=1}^3 n_i\big)}$, typically arising from discretized operators, the canonical format is usually given in \textit{Kronecker product form} \cite{tyrtyshnikov_2003}
\begin{align}\label{kronecker_form}
A = \sum_{j=1}^r \alpha_j\, U_j^{(1)} \otimes U_j^{(2)} \otimes U_j^{(3)},
\end{align}
with matrices $U_j^{(q)} \in \mathbb{R}^{n_q \times n_q}$, scalars $\alpha_j$ and $\otimes$ equals \textit{Kronecker product}. Due to the relation $\text{vec}\big(\text{vec}(U^{(3)}) \circ \text{vec}(U^{(2)}) \circ \text{vec}(U^{(1)})\big) = \text{vec}(U^{(1)}) \otimes \text{vec}(U^{(2)}) \otimes \text{vec}(U^{(3)})$, the form \eqref{kronecker_form} can be identified with \eqref{CP}, where the 
\textit{vectorization} vec(.) is understood as in \cite{kolda_tensor_2009}.\\
Storage and tensor operations for the canonical format scale linearly in the dimension $d$, rather than exponentially as for dense tensors. However, the above mentioned drawbacks (instability, lack of robust algorithms) have led to the development of other (stable) formats that scale linearly in the dimension, such as $\mathcal{H}$\textit{-Tucker} \cite{grasedyck_htucker_2010} which relies on hierarchical tree structure, or the \textit{Tensor Train format} \cite{oseledets_2011}, which briefly discussed in Sec.~\ref{TTtensor}.
\subsection{Tucker tensors and quasi-best approximation}\label{best_approx}
For a matrix $U \in \mathbb{R}^{r \times n_j}$,
the \textit{$j$-mode matrix product} $ \mathcal{X} \times_j U $ of a tensor $\mathcal{X} \in \bigotimes_{p=1}^3\mathbb{R}^{n_p}$ with $ U $ is
defined element-wise in the following way. E.g.\ for $ j=1 $,
\begin{align} \label{modemul}
(\mathcal{X} \times_1 U)_{i_1\,i_2\,i_3} := \sum_{i^{\prime}=1}^{n_1} x_{i^{\prime}\,i_2\,i_3}\,u_{i_1\,i^{\prime}},
\end{align}
i.e., the resulting tensor  $\mathcal{X} \times_1 U \in \bigotimes_{p=1}^3\mathbb{R}^{m_p}$, where $m_p = n_p,\, p\neq1$ and $m_1 = r$, is obtained
by right-multiplication of the $1$-mode fibers (columns) of $ \mathcal{X} $ by $ U $.
Analogously for $ j=2,3 $; the cost for the computation of $ \mathcal{X} \times_j U $ is
$O(r \prod_{j=1}^{3} n_j)$ operations in general.
A Tensor $\mathcal{X}\in \bigotimes_{p=1}^3\mathbb{R}^{n_p}$ is said to be represented in \textit{Tucker format} if 
\begin{align}\label{tucker_rep}
\mathcal{X} = \mathcal{C}\times_1 \,U^{(1)} \times_2 U^{(2)} \times_3 U^{(3)} \equiv \llbracket\, \mathcal{C};\,U^{(1)},U^{(2)},U^{(3)}\,\rrbracket,
\end{align}
with the so-called \textit{core tensor} $\mathcal{C} \in \bigotimes_{p=1}^3\mathbb{R}^{r_p}$
and (factor) matrices $U^{(j)} \in \mathbb{R}^{n_j \times r_j}$. The storage requirement for \eqref{tucker_rep} is $\prod_{j=1}^3 r_j + \sum_{j=1}^3 n_jr_j$, which is smaller than $\prod_{j=1}^3 n_j$ if $r_j \ll n_j$.\\
The $j$\textit{-rank} of a tensor $\mathcal{X}$ is the rank of the unfolding ($j$-mode matricization) $\mathcal{X}_{(j)}$, \cite{kolda_tensor_2009}. By setting $r_j = \text{rank}(\mathcal{X}_{(j)})$, the tensor $\mathcal{X}$ is usually referred to as \textit{rank-$(r_1,r_2,r_3)$ tensor}. 
Of course $r_j \leq n_j$ holds.\\
In the following we denote the set of Tucker tensors with mode sizes $\boldsymbol{n} = (n_1,n_2,n_3)$ and ($j$-) ranks $\boldsymbol{r} = (r_1,r_2,r_3)$  with $\mathcal{T}_{\boldsymbol{n},\boldsymbol{r}}$. In fact, $\mathcal{T}_{\boldsymbol{n},\boldsymbol{r}}$ contains all tensors with mode-size $\boldsymbol{n}$ and $j$-ranks smaller or equal $r_j$, \cite{hackbusch_tensor_2012}.\\
The set $\mathcal{T}_{\boldsymbol{n},\boldsymbol{r}}$ is closed \footnote{This also holds for order-$d$ tensors where $d > 3$ and can be proved along the same lines.} due to the fact that the set of matrices of rank $\leq r$ is closed, i.e. for a sequence of matrices $(U_n)_{n\in\mathbb{N}}$ with ranks $\leq r$ we have $\lim_{n\rightarrow \infty} U_n =: U$ has rank $\leq r$. Thus, if we assume a sequence $(\mathcal{X}_n)_{n\in \mathbb{N}} \subset \mathcal{T}_{\boldsymbol{n},\boldsymbol{r}} \subset \bigotimes_{p=1}^3\mathbb{R}^{n_p}$, we get $\lim_{n\rightarrow \infty} \text{rank}({\mathcal{X}_n}_{(j)}) \leq r_j$.\\     
In finite dimension, the closedness of the subset $\mathcal{T}_{\boldsymbol{n},\boldsymbol{r}}$ of a normed vector space, e.g. $(\bigotimes_{p=1}^3\mathbb{R}^{n_p},\left\|\,.\,\right\|_F)$, ensures the existence of a \textit{best-approximation} $\mathcal{X}_{\text{best}}\in\mathcal{T}_{\boldsymbol{n},\boldsymbol{r}}$ of an element in $\mathcal{X}\in\bigotimes_{p=1}^3\mathbb{R}^{n_p}$, i.e.
\begin{align}
\left\|\,\mathcal{X}-\mathcal{X}_{\text{best}}\,\right\|_F \leq \left\|\,\mathcal{X}-\mathcal{Y}\,\right\|_F \quad \forall \mathcal{Y} \in \mathcal{T}_{\boldsymbol{n},\boldsymbol{r}}, 
\end{align}  
see Sec.~\ref{sec_best_appr} Lemma~\ref{best_approx_lemma}.\\
An approximation of a given tensor to prescribed $j$-ranks $\boldsymbol{r}$ was investigated in \cite{lathauwer_svd}, where the described algorithm to compute such an approximation (Higher Order Singular Value Decomposition [HOSVD]) works by truncating the SVD of the $j$-mode unfoldings. The resulting tensor is an approximation to the best-approximation in $\mathcal{T}_{\boldsymbol{n},\boldsymbol{r}}$. Indeed, due to Property $10$ in \cite{lathauwer_svd} we have
for the HOSVD approximation $\mathcal{X}_{\text{HOSVD}}$ of a tensor $\mathcal{X}$ with $j$-ranks $R_j$ and the descending ordered singular values of the $j$-th unfolding of $\mathcal{X}$ denoted with $\sigma_{k}^{(j)},\, k=1\hdots R_j$ (here formulated for order-$3$ tensors)
\begin{align}\label{quasi_best_tucker}
\left\|\,\mathcal{X}-\mathcal{X}_{\text{HOSVD}}\,\right\|_F \leq \sqrt{\sum_{j = 1}^3 \sum_{k=r_j+1}^{R_j}  {\sigma_{k}^{(j)}}^2 } \leq \sqrt{3} \left\|\,\mathcal{X} - \mathcal{X}_{\text{best}}\,\right\|_F,
\end{align}  
where $\mathcal{X}_{\text{best}}$ is the best approximation of $\mathcal{X}$ in $\mathcal{T}_{\boldsymbol{n},\boldsymbol{r}}$, cf. \cite{grasedyck_htucker_2010}.\\
Existence of the Tucker approximation was also investigated in \cite{kroonenberg}, as well as alternating least squares methods (ALS) for the fitting problem
\begin{align}\label{Tucker_fitting}
\min_{\mathcal{C},U^{(1)},U^{(2)},U^{(3)}}\,\,\left\|\mathcal{X} - \llbracket\, \mathcal{C};\,U^{(1)},U^{(2)},U^{(3)}\,\rrbracket\right\|_F^2 \quad \text{s.t.} \quad U^{(p)} \,\,\, \text{columnwise orthonormal,}
\end{align}
where $\mathcal{X}\in \bigotimes_{p=1}^3\mathbb{R}^{n_p}$ is given and $\mathcal{C}\in \bigotimes_{p=1}^3\mathbb{R}^{r_p},\, U^{(p)} \in \mathbb{R}^{n_p \times r_p}$ to be computed.\\
Another algorithm is the so-called \textit{higher order orthogonal iteration} (HOOI), an ALS algorithm, \cite{kolda_tensor_2009}, which can also be used for the purpose of \textit{recompression} (namely computing a quasi-optimal Tucker approximation of lower rank to a given Tucker tensor). Algorithmic variants of the HOOI were investigated in \cite{andersson}.\\
Using the orthonormality of the factor matrices and rewriting the objective in \eqref{Tucker_fitting} gives 
\begin{align}\label{Tucker_fitting2}
\left\|\mathcal{X} - \llbracket\, \mathcal{C};\,U^{(1)},U^{(2)},U^{(3)}\,\rrbracket\right\|_F^2 = \left\|\mathcal{X} \right\|_F^2 + \left\|\mathcal{C} \right\|_F^2 - 2\left\langle \mathcal{X} \times_1 {U^{(1)}}^T \times_2 {U^{(2)}}^T \times_3 {U^{(3)}}^T \,,\, \mathcal{C}\right\rangle. 
\end{align}
Its gradient w.r.t. to $\mathcal{C}$ is given as
\begin{align}
\partial_{\mathcal{C}} \left\|\mathcal{X} - \llbracket\, \mathcal{C};\,U^{(1)},U^{(2)},U^{(3)}\,\rrbracket\right\|_F^2 = 2(\mathcal{C} - \mathcal{X} \times_1 {U^{(1)}}^T \times_2 {U^{(2)}}^T \times_3 {U^{(3)}}^T),
\end{align} which attains zero for 
\begin{align}\label{core_prop}
\mathcal{C} = \mathcal{X} \times_1 {U^{(1)}}^T \times_2 {U^{(2)}}^T \times_3 {U^{(3)}}^T,
\end{align}
giving a necessary condition for the optimal choice of the core.\\
By inserting \eqref{core_prop} into \eqref{Tucker_fitting2} problem \eqref{Tucker_fitting} can be recasted as a maximization problem \cite{kolda_tensor_2009}, \cite{lathauwer}, i.e.
\begin{align}\label{Tucker_fitting3}
\max_{U^{(1)},U^{(2)},U^{(3)}}\,\,\left\|\mathcal{X} \times_1 {U^{(1)}}^T \times_2 {U^{(2)}}^T \times_3 {U^{(3)}}^T\right\|_F^2 \quad \text{s.t.} \quad U^{(p)} \,\,\, \text{columnwise orthonormal.}
\end{align}
An alternating least squares approach for solving \eqref{Tucker_fitting3} can easily be derived by alternately fixing all but one factor matrix and solving for the remaining by an SVD-approach, see \cite{kolda_tensor_2009} and  Alg.\ref{HOOI}.
\begin{algorithm}
\caption{HOOI (ALS);\, \texttt{HOOI($\mathcal{X}$)}}
\label{HOOI}
\begin{algorithmic}[1]
\Require $\mathcal{X}\in \bigotimes_{p=1}^3\mathbb{R}^{n_p},\,\,\epsilon > 0,\,\, r_{p,0} \in \mathbb{N}_{>1},\, p = 1\hdots 3$\,\,\text{(initial guesses for $j$-ranks)}
\Ensure $\mathcal{C} \in \bigotimes_{p=1}^3\mathbb{R}^{r_p}, \,\,U^{(p)}\in \mathbb{R}^{n_p \times r_p},\, p = 1\hdots 3$
\State \text{Initialize}\,\, $U^{(p)}\in \mathbb{R}^{n_p \times r_{p_{0}}},\,\, p = 1\hdots 3$ e.g. by HOSVD or random
\Repeat
\For{$p = 1 \hdots 3$}
  \State $\mathcal{Y} \gets \mathcal{X} \times_{\substack{{j =1}\\j \neq p}}^{3}  {U^{(j)}}^T$ 
  \State $(U,\Sigma) \gets \text{svd}(\mathcal{Y}_{(p)}) \quad (\text{where}\,\, \mathbb{R}^{n_p \times \prod_{i\neq p} r_i} \ni \mathcal{Y}_{(p)} = U\Sigma V)$
  \State $r_p \gets \min\{r \,\mid \,\left\| \Sigma(1:r,1:r)\right\|_F = \sqrt{\sum_{i=1}^r \sigma_i^2} < \frac{\epsilon}{\sqrt{3}} \left\| \Sigma \right\|_F\}$
  \State $U^{(p)} \gets U(:,1:r_p)$ 
\EndFor
\Until{\text{fit ceases to improve}}
\State $\mathcal{C} \gets \mathcal{X} \times_1 {U^{(1)}}^T \times_2 {U^{(2)}}^T \times_3 {U^{(3)}}^T$
\end{algorithmic}
\end{algorithm}
In the description of Alg.\ref{HOOI} we assume the (common) convention of descending ordered singular values. It is enough to compute the so-called \textit{economic sized} svd, namely, in the case $\prod_{i\neq p} r_i < n_p$ 
only the first $\prod_{i\neq p} r_i$ columns of $U$ have to be computed and $\Sigma \in \mathbb{R}^{\prod_{i\neq p} r_i \times \prod_{i\neq p} r_i}$; analog for the case $\prod_{i\neq p} r_i \geq n_p$. \\
The SVD is truncated in such way that the relative error in the Frobenius norm is smaller than the given tolerance of $\epsilon/\sqrt{3}$. 
HOOI converges to a solution where the objective function of \eqref{Tucker_fitting2} ceases to decrease; in fact, the convergence of HOOI to a global optimum, not even to stationary points, is not guaranteed \cite{lathauwer},\cite{kroonenberg}. Nevertheless, to our experience Alg.\ref{HOOI} works well in practice and mostly yields better results than HOSVD (even for random initialization).
An efficient generalization of Alg.\ref{HOOI} to recompression, i.e. the case where $\mathcal{X}$ is already in Tucker form, is straight forward.\\
HOOI can also be used for approximate addition of Tucker tensors. Assume $\mathcal{X}$ is a sum of two Tucker tensors with equal mode sizes (Block CP [BCP], \cite{kolda_tensor_2009},\cite{hopke_2011}), i.e.
\begin{align}
\mathcal{X} = \llbracket\, \mathcal{D};\,V^{(1)},V^{(2)},V^{(3)}\,\rrbracket + \llbracket\, \mathcal{E};\,W^{(1)},W^{(2)},W^{(3)} \,\rrbracket.
\end{align}
Mode multiplications and matricization have to be performed with respect to the BCP format, which can be done elementwise (summand by summand). 
\section{Kronecker product approximation for long-range interactions}\label{sec_kron}
We will derive a \textit{Kronecker product approximation} (cf. \eqref{kronecker_form}) for the potential operator \eqref{intop2} by certain quadrature of an integral representation of the convolution kernel.\\
In order to motivate the strategy, let us assume a multivariate function $f = f(\rho): \, \mathbb{R}^d \rightarrow \mathbb{R}$ where $\rho = \left\|\boldsymbol{x}\right\|^2 \in [a,b] \subset \mathbb{R}$ and the integral representation
\begin{align}\label{intrep1}
f(\rho) = \int_{\mathbb{R}} g(\tau)\,e^{\rho h(\tau)} \,\text{d}\tau.
\end{align}
If we can apply quadrature to \eqref{intrep1} with nodes and weights $(t_k,\omega_k),\, k=1\hdots R$, we obtain a separable representation of $f$, i.e.
\begin{align}\label{seprep}
f(\rho) \approx \sum_{k=1}^R \, \omega_k\,g(t_k)\,e^{\rho h(t_k)} = \sum_{k=1}^R \, \omega_k\,g(t_k)\,\prod_{p=1}^d e^{{x^{(p)}}^2 h(t_k)}.
\end{align}
The quadrature order $R$ refers to the \textit{separation rank} of \eqref{seprep}.\\
In the following we derive a separable representation for the convolution operator in \eqref{intop2} after applying a collocation scheme. This representation leads to the desired Kronecker product form of the operator, see Corr.~\ref{kron_op}.
\subsection{Notation}
Let $\Omega = \varprod_{p=1}^3 \Omega^{(p)} \subset \mathbb{R}^3$ with $\Omega^{(p)} = [\alpha_p,\beta_p] \subset \mathbb{R}$ and assume for $p=1\hdots 3$ a partition of $\Omega^{(p)}$ into $n_p$ sub-intervals $I_i^{(p)},\, i= 1\hdots n_p$. On the resulting tensor grid $T := W^{(1)} \times W^{(2)} \times W^{(3)}$ of $\Omega$ where $W^{(p)} = \varprod_{i=1}^{n_p} I_i^{(p)}$ we define sets of collocation points $\{\xi_i^{(p)} \in I_i^{(p)}: \, i=1 \hdots n_p\},\, p=1\hdots 3$ (for ease of presentation, one collocation point per sub-interval, e.g. midpoints).\\
We further denote the number of collocation points $\boldsymbol{\xi}_{\boldsymbol{i}} = (\xi_{i_1}^{(1)},\xi_{i_2}^{(2)},\xi_{i_3}^{(3)}),\, \boldsymbol{i} = (i_1,i_2,i_3)$ with $N := \prod_p n_p$.
\subsection{The collocation scheme}
\begin{figure}[hbtp]
\center
\includegraphics[scale = 0.4]{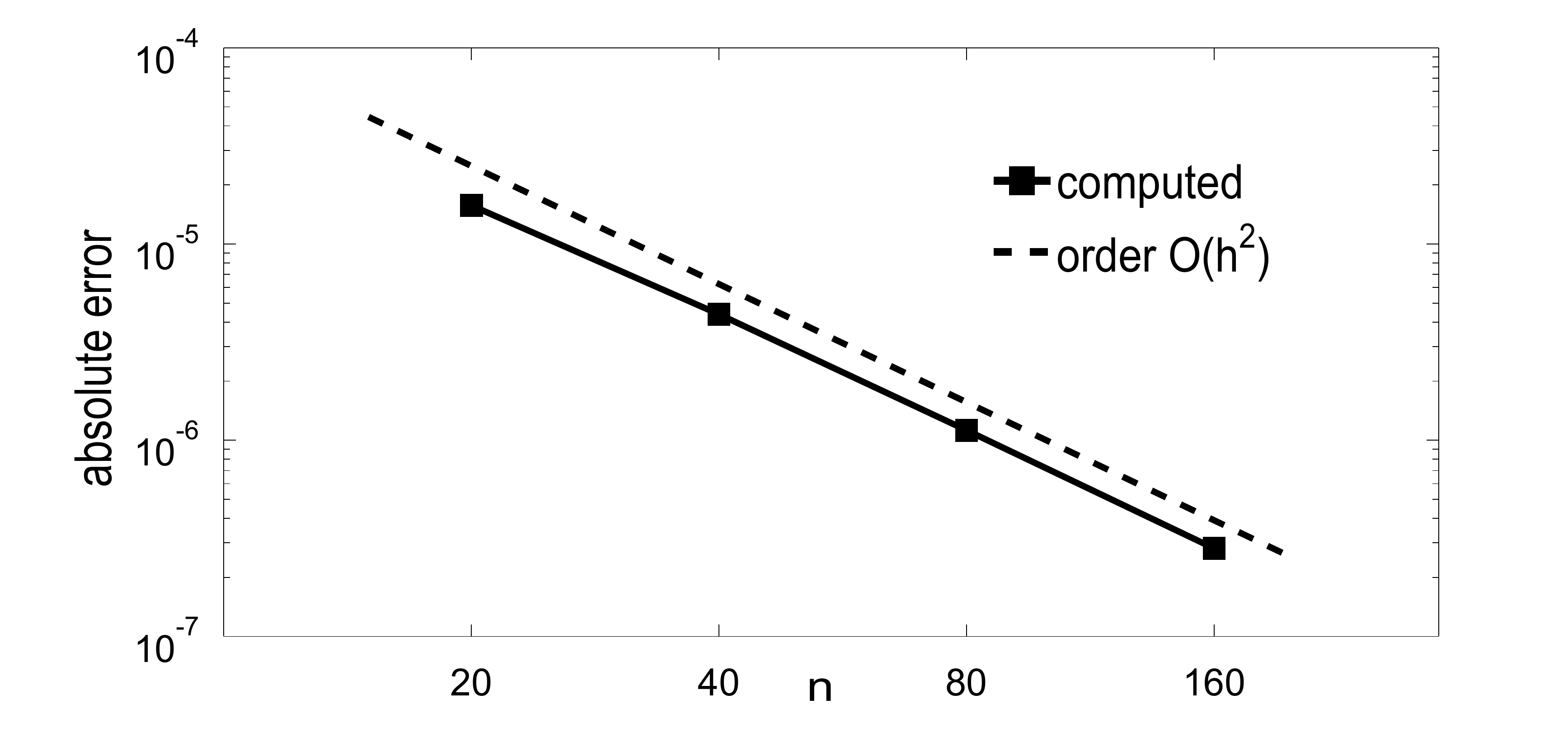}    
\caption{Absolute error of collocation scheme \eqref{coll}.}\label{fig_err}
\end{figure}

Using the tensor product basis functions (e.g. $\psi_{i}^{(p)} := \chi_{I_{i}^{(p)}}$ indicator function of sub-interval $I_{i}^{(p)}$) 
\begin{align}\label{ansatz_fct}
\psi_{i_1 i_2 i_3}(\boldsymbol{x}) = \prod_{p=1}^3 \psi_{i_p}^{(p)}(x^{(p)}),
\end{align} 
and the ansatz cf. \cite{exl_fast_2012_2} ($m_{\boldsymbol{j}}^{(q)} = m^{(q)}(\boldsymbol{\xi}_{\boldsymbol{j}}),\, \boldsymbol{j} = (j_1,j_2,j_3)$)
\begin{align}
\boldsymbol{m}^{(q)}(\boldsymbol{x}) = \sum_{\boldsymbol{j}} m_{\boldsymbol{j}}^{(q)} \psi_{\boldsymbol{j}}(\boldsymbol{x})
\end{align}
our collocation scheme for \eqref{intop2} takes the form ($\boldsymbol{i} = (i_1,i_2,i_3),\, \boldsymbol{j} = (j_1,j_2,j_3)$)
\begin{align}\label{coll}
\phi_{\boldsymbol{i}} = -\frac{1}{4 \pi} \sum_{q=1}^3 \sum_{\boldsymbol{j}} m_{\boldsymbol{j}}^{(q)} \int_\Omega g^{(q)}(\boldsymbol{\xi}_{\boldsymbol{i}},\boldsymbol{y}) \, \psi_{\boldsymbol{j}}(\boldsymbol{y})\, \text{d} \boldsymbol{y},
\end{align} where 
\begin{align}\label{kernel}
g^{(q)}(\boldsymbol{x},\boldsymbol{y}) := \frac{x^{(q)} - y^{(q)}}{\left\|\boldsymbol{x} - \boldsymbol{y}\right\|^3}.
\end{align}
\begin{lem}\label{conv_coll}
Let $\boldsymbol{m}\in C^2(\Omega)$. Then the collocation scheme \eqref{coll}, where $\xi_i^{(p)}$ are the midpoints of $I_i^{(p)}\, i= 1\hdots n_p$, converges quadratically. 
\end{lem}
\begin{pf}
Let us assume (w.l.o.g) uniform spacings in each dimension, i.e. $h_p := 1/n_p$, and use the notation $h := \max_{p = 1\hdots 3} h_p$. Furthermore, let $\Omega_{\boldsymbol{j}} := \varprod_{p=1}^3 I_{j_p}^{(p)}$.\\
We estimate the local error $e(\boldsymbol{\xi}_{\boldsymbol{i}}) = |\phi_{\boldsymbol{i}} - \phi(\boldsymbol{\xi}_{\boldsymbol{i}}) |$ (cf. \eqref{coll} and \eqref{intop2}) for each fixed $\boldsymbol{i}$ and $q$ in \eqref{coll} separately, i.e. we define
\begin{align}
e_q(\boldsymbol{\xi}_{\boldsymbol{i}}) := \left| \sum_{\boldsymbol{j}} m_{\boldsymbol{j}}^{(q)} \int_\Omega g^{(q)}(\boldsymbol{\xi}_{\boldsymbol{i}},\boldsymbol{y}) \, \psi_{\boldsymbol{j}}(\boldsymbol{y})\, \text{d} \boldsymbol{y} - 
\sum_{\boldsymbol{j}} \int_\Omega m^{(q)}(\boldsymbol{y})\, g^{(q)}(\boldsymbol{\xi}_{\boldsymbol{i}},\boldsymbol{y}) \, \psi_{\boldsymbol{j}}(\boldsymbol{y})\, \text{d} \boldsymbol{y}\right|\\[0.1cm]
 = \left| \sum_{\boldsymbol{j}} \int_\Omega \big(m_{\boldsymbol{j}}^{(q)} - m^{(q)}(\boldsymbol{y})\big)\, g^{(q)}(\boldsymbol{\xi}_{\boldsymbol{i}},\boldsymbol{y}) \, \psi_{\boldsymbol{j}}(\boldsymbol{y})\, \text{d} \boldsymbol{y}\right|. 
\end{align}

By using Taylor expansion for $m^{(q)}$ at 'source' points $\boldsymbol{\xi}_{\boldsymbol{j}}$, i.e.
\begin{align}
m^{(q)}(\boldsymbol{y}) = m^{(q)}(\boldsymbol{\xi}_{\boldsymbol{j}}) + \left\langle \nabla m^{(q)}(\boldsymbol{\xi}_{\boldsymbol{j}}),\boldsymbol{y} - \boldsymbol{\xi}_{\boldsymbol{j}}\right\rangle + \mathcal{O}(\left\|\boldsymbol{y} - \boldsymbol{\xi}_{\boldsymbol{j}}\right\|^2),
\end{align}
%
%
we obtain ($C_1 > 0$ independent of $h$)
\begin{align}\label{loc_err}
e_q(\boldsymbol{\xi}_{\boldsymbol{i}}) \leq \sum_{\boldsymbol{j}} \left|\int_{\Omega} \left\langle \nabla m^{(q)}(\boldsymbol{\xi}_{\boldsymbol{j}}),\boldsymbol{y} - \boldsymbol{\xi}_{\boldsymbol{j}}\right\rangle g^{(q)}(\boldsymbol{\xi}_{\boldsymbol{i}},\boldsymbol{y}) \, \psi_{\boldsymbol{j}}(\boldsymbol{y}) \, \text{d} \boldsymbol{y}\right| +
C_1  \left|\int_{\Omega} \left\|\boldsymbol{y} - \boldsymbol{\xi}_{\boldsymbol{j}}\right\|^2 g^{(q)}(\boldsymbol{\xi}_{\boldsymbol{i}},\boldsymbol{y}) \, \psi_{\boldsymbol{j}}(\boldsymbol{y}) \, \text{d} \boldsymbol{y}\right|. 
\end{align}
It can easily be seen that $g^{(q)}(\boldsymbol{\xi}_{\boldsymbol{i}},.) \in L^p(\Omega_{\boldsymbol{j}})$ for all $\boldsymbol{j}$ and $1 \leq p < 3/2$.\\
Hence, the second term in \eqref{loc_err} allows the estimate
\begin{align}
\sum_{\boldsymbol{j}} \left|\int_{\Omega} \left\|\boldsymbol{y} - \boldsymbol{\xi}_{\boldsymbol{j}}\right\|^2 g^{(q)}(\boldsymbol{\xi}_{\boldsymbol{i}},\boldsymbol{y}) \, \psi_{\boldsymbol{j}}(\boldsymbol{y}) \, \text{d} \boldsymbol{y}\right| \leq \sum_{\boldsymbol{j}} \left\|g^{(q)}(\boldsymbol{\xi}_{\boldsymbol{i}},.) \right\|_{L^1(\Omega_{\boldsymbol{j}})} \int_{\Omega} \left\|\boldsymbol{y} - \boldsymbol{\xi}_{\boldsymbol{j}}\right\|^2 \, \psi_{\boldsymbol{j}}(\boldsymbol{y}) \, \text{d} \boldsymbol{y} \\[0.1cm]
\leq \left\|g^{(q)}(\boldsymbol{\xi}_{\boldsymbol{i}},.) \right\|_{L^1(\Omega )}\,\sum_{\boldsymbol{j}}\,\int_{\Omega_{\boldsymbol{j}}} \left\|\boldsymbol{y} - \boldsymbol{\xi}_{\boldsymbol{j}}\right\|^2 \, \text{d} \boldsymbol{y} \leq \left\|g^{(q)}(\boldsymbol{\xi}_{\boldsymbol{i}},.) \right\|_{L^1(\Omega )}\,|\Omega|\, h^2 = \mathcal{O}(h^2).
\end{align}

For the first term we distinguish between the diagonal ($\boldsymbol{i} = \boldsymbol{j}$) and non-diagonal ($\boldsymbol{i} \neq \boldsymbol{j}$) case. 
The kernel $g^{(q)}(\boldsymbol{\xi}_{\boldsymbol{i}},.)$ is analytic in the case $\boldsymbol{i} \neq \boldsymbol{j}$ and thus allows Taylor expansion, i.e.
\begin{align}\label{taylor_kern}
g^{(q)}(\boldsymbol{\xi}_{\boldsymbol{i}},\boldsymbol{y}) = g^{(q)}(\boldsymbol{\xi}_{\boldsymbol{i}},\boldsymbol{\xi}_{\boldsymbol{j}}) + \mathcal{O}(\left\| \boldsymbol{y} -\boldsymbol{\xi}_{\boldsymbol{j}}\right\|).
\end{align}
The constant term in the expansion does not contribute to the first term in \eqref{loc_err} since $\boldsymbol{y} -\boldsymbol{\xi}_{\boldsymbol{j}}$ is odd w.r.t. $\boldsymbol{\xi}_{\boldsymbol{j}}$ in $\Omega_{\boldsymbol{j}}$. We get ($C_2 > 0$ independent of $h$)
\begin{align}
\sum_{\boldsymbol{j} \neq \boldsymbol{i}} \left|\int_{\Omega} \left\langle \nabla m^{(q)}(\boldsymbol{\xi}_{\boldsymbol{j}}),\boldsymbol{y} - \boldsymbol{\xi}_{\boldsymbol{j}}\right\rangle g^{(q)}(\boldsymbol{\xi}_{\boldsymbol{i}},\boldsymbol{y}) \, \psi_{\boldsymbol{j}}(\boldsymbol{y}) \, \text{d} \boldsymbol{y}\right| \leq 
\left\|\nabla m^{(q)}(\boldsymbol{\xi}_{\boldsymbol{j}})\right\|\, \sum_{\boldsymbol{j} \neq \boldsymbol{i}} \int_{\Omega_{\boldsymbol{j}}} \mathcal{O}(\left\| \boldsymbol{y} -\boldsymbol{\xi}_{\boldsymbol{j}}\right\|^2) \leq C_2 \,|\Omega|\, h^2 = \mathcal{O}(h^2).
\end{align}
In the case $\boldsymbol{i} = \boldsymbol{j}$ we have for the first term in \eqref{loc_err} 
\begin{align}
\left|\int_{\Omega} \left\langle \nabla m^{(q)}(\boldsymbol{\xi}_{\boldsymbol{i}}),\boldsymbol{y} - \boldsymbol{\xi}_{\boldsymbol{i}}\right\rangle \,g^{(q)}(\boldsymbol{\xi}_{\boldsymbol{i}},\boldsymbol{y}) \, \psi_{\boldsymbol{i}}(\boldsymbol{y}) \, \text{d} \boldsymbol{y} \right|\leq 
\left\| \nabla m^{(q)}(\boldsymbol{\xi}_{\boldsymbol{i}}) \right\|\, \int_{\Omega} \left\|\boldsymbol{y} - \boldsymbol{\xi}_{\boldsymbol{i}}\right\| \,\left|g^{(q)}(\boldsymbol{\xi}_{\boldsymbol{i}},\boldsymbol{y})\right| \, \psi_{\boldsymbol{i}}(\boldsymbol{y})\, \text{d}\boldsymbol{y} \leq C_2\, (I_1 f)(\boldsymbol{\xi}_{\boldsymbol{i}}),
\end{align} 
where $(I_1 f)(\boldsymbol{\xi}_{\boldsymbol{i}}) := \int_{\mathbb{R}^3} \frac{\left|f(\boldsymbol{y})\right|}{\left\|\boldsymbol{\xi}_{\boldsymbol{i}} - \boldsymbol{y}\right\|^2}\, \text{d}\boldsymbol{y}$ with $f(\boldsymbol{y}) = (\xi_{\boldsymbol{i}}^{(q)} - y^{(q)})\psi_{\boldsymbol{i}}(\boldsymbol{y}) \in L^p(\Omega),\,p\geq1$ with compact support $\Omega_{\boldsymbol{i}}$. Since $1/\left\|\boldsymbol{x} \right\|^2 \in L^p(\Omega),\,1\leq p < 3/2$, we proceed with 
\begin{align}
(I_1 f)(\boldsymbol{\xi}_{\boldsymbol{i}}) \leq \int_{\Omega_i} \frac{h_q}{\left\| \boldsymbol{\xi}_{\boldsymbol{i}} - \boldsymbol{y} \right\|^2} \, \text{d}\boldsymbol{y} \leq h\, \int_{B_{h}(\boldsymbol{0})} \frac{1}{\left\|\boldsymbol{x} \right\|^2} \, \text{d}\boldsymbol{x} = 4 \pi\, h^2,
\end{align}
%
%
which completes the proof.
\hfill $\square$
\end{pf}
Fig.~\ref{fig_err} shows the quadratic convergence of the error of the collocation scheme \eqref{coll} compared to \eqref{intop2} evaluated at the origin\footnote{We used Maple $14$ and evaluated the expression \eqref{intop2} at the points $\boldsymbol{\xi}_{\boldsymbol{i}} = 1/2(h,h,h)$ with $h=1/n$.} for the radial symmetric function $m(\boldsymbol{x}) = \exp(-\left\|\boldsymbol{x}\right\|^2)$.\\ 
The evaluation of the potential $\phi$ on $T$ by abbreviating notation for the matrices $\boldsymbol{G}^{(q)} \in \mathbb{R}^{N \times N}$ with entries
\begin{align}\label{matrix_op}
G_{\boldsymbol{i}\boldsymbol{j}}^{(q)} = \int_\Omega g^{(q)}(\boldsymbol{\xi}_{\boldsymbol{i}},\boldsymbol{y}) \, \psi_{\boldsymbol{j}}(\boldsymbol{y})\, \text{d} \boldsymbol{y}
\end{align} 
and the grid-sampled magnetization components $\boldsymbol{m}^{(q)} \in \mathbb{R}^N$ is given as
\begin{align}\label{approx_phi}
\phi \approx -\frac{1}{4\pi} \sum_{q=1}^3 \boldsymbol{G}^{(q)}\,\boldsymbol{m}^{(q)},
\end{align}
yielding computational effort $\mathcal{O}(N^2)$ if no special structure on $\boldsymbol{G}^{(q)}$ can be imposed.\\
We will use the term \textit{(potential) operator} for 
\begin{align}\label{pot_op}
P: \mathbb{R}^N \times \mathbb{R}^N \times \mathbb{R}^N \rightarrow \mathbb{R}^N,\, (\boldsymbol{m}^{(1)},\boldsymbol{m}^{(2)},\boldsymbol{m}^{(3)}) \mapsto -\frac{1}{4\pi} \sum_{q=1}^3 \boldsymbol{G}^{(q)}\,\boldsymbol{m}^{(q)}.
\end{align}
In the following we use \textit{Gauss-Transform} (cf. \cite{exl_fast_2012_2}) and \textit{sinc-quadrature} \cite{steger_1993} to construct a Kronecker product structure for \eqref{pot_op} with an asymptotically optimal approximation error yielding a 'tensor' version $\mathcal{P}$ of \eqref{pot_op} that operates on structured tensors.

\subsection{Sinc quadrature}
We state some basic facts from the theory of sinc function based approximation, see \cite{steger_1993},\cite{hackbusch_low-rank_2005}.\\
The sinc function $\text{sinc}(x) := \frac{\sin(\pi x)}{\pi x}$ is an analytic function which is $1$ at $x=0$ 
and zero at $x \in \mathbb{Z}\setminus\{0\}$. Sufficiently fast decaying continuous functions $f \in C(\mathbb{R})$ can be interpolated at the grid points $x = k\vartheta\in \vartheta\mathbb{Z},\, \vartheta>0$ (step size) by functions $S_{k,\vartheta}(x) := \text{sinc}(x/\vartheta - k)$, i.e
\begin{align}
f_\vartheta(x) = \sum_{k\in \mathbb{Z}} f(k\vartheta)S_{k,\vartheta}(x).
\end{align}
Since $\int_{\mathbb{R}} \text{sinc}(t)\,\text{d}t = 1$, the interpolatory quadrature for $\int_{\mathbb{R}} f(t) \, \text{d}t$
leads to
\begin{align}\label{sinc_int}
\int_{\mathbb{R}} f(t) \, \text{d}t \approx \vartheta \sum_{k\in \mathbb{Z}} f(k\vartheta),
\end{align}
which can be seen as infinite trapezional rule. Truncation to $k = -R \hdots R$ of the infinite sum in \eqref{sinc_int} leads to the \textit{sinc quadrature rule} with $2R+1$ terms with the truncation error $\vartheta \sum_{|k|>R} f(k\vartheta)$ that obviously depends on the decay-rate of $f$ on the real axis.\\
For functions $f \in H^1(D_\delta),\, \delta < \pi/2$ (\textit{Hardy space}), i.e. which are holomorphic in the strip $D_\delta := \{z\in \mathbb{C}:\,|\Im\,z| \leq \delta\}$ with 
\begin{align}
N(f,D_\delta) := \int_{\partial D_{\delta}} |f(z)| \, |dz| = \int_{\mathbb{R}} \big(|f(t + i\delta)| + |f(t - i\delta)| \big)\, \text{d}\,t < \infty,
\end{align} 
and in addition to $f \in H^1(D_\delta)$ have double exponential decay on the real axis, the following exponential error estimate for the sinc quadrature holds (cf. \cite{hackbusch_low-rank_2005}, Proposition 2.1), which we state for sake of completeness.
\begin{thm}[\cite{hackbusch_low-rank_2005}]\label{sinc_theo}
Let $f \in H^1(D_\delta)$ with some $\delta < \pi/2$. If $f$ satisfies the condition 
\begin{align}\label{double_exp}
|f(t)| \leq C\,\exp(-be^{a|t|}) \quad \forall t\in \mathbb{R} \,\, with\,\, a,b,C > 0,
\end{align}
then the quadrature error for the special choice $\vartheta = \log(\frac{2\pi a R}{b})/(aR)$ satisifies
\begin{align}\label{quad_err}
\left|\int_{\mathbb{R}} f(t) \, \text{d}t - \vartheta \sum_{|k|\leq R} f(k\vartheta)\right| \leq C\, N(f,D_\delta)\, \exp\Big(\frac{-2\pi \delta a R}{\log(2\pi aR/b)}\Big).
\end{align}
\end{thm}
\begin{rmk}
In the case $f(\rho)$ as in \eqref{intrep1} the constants in \eqref{quad_err} depend on $\rho$. For some fixed $\rho$, an accuracy of $\epsilon > 0$ can be achieved with $R = \mathcal{O}(|\log\epsilon|\cdot \log|\log\epsilon|)$.
\end{rmk}
\subsection{Separable approximation of $\boldsymbol{G}^{(q)}$}
We make use of the Gaussian transform
\begin{align}\label{gauss_trans}
\frac{1}{\rho^{3/2}} = \frac{2}{\sqrt{\pi}} \int_{\mathbb{R}} \tau^2 \, e^{-\tau^2\rho}\,\text{d}\tau,
\end{align}
to obtain for $\rho = \left\|\boldsymbol{\xi}_{\boldsymbol{i}} - \boldsymbol{y}\right\|^2$ and $q = 1\hdots 3$ the new representation for \eqref{matrix_op}
\begin{align}\label{matrix_op_sep}
\boldsymbol{G}_{\boldsymbol{i}\boldsymbol{j}}^{(q)} = \frac{2}{\sqrt{\pi}} \int_{\mathbb{R}} \tau^2 \prod_{p=1}^3 h_{i_p j_p}^{(p)}(\tau) \,\text{d}\tau \equiv \frac{2}{\sqrt{\pi}} \int_{\mathbb{R}} f(\tau)\,\text{d}\tau,
\end{align}
with 
\begin{align}\label{sep_integrands}
h_{i_p j_p}^{(p)}(\tau) = 
\left\{\begin{array}{l l}
\int_{I_{j_p}^{(p)}} e^{-\tau^2 (\xi_{i_p} - y)^2}\,\text{d}y & p \neq q, \\*[\jot]
\int_{I_{j_p}^{(p)}} (\xi_{i_p} - y)\,e^{-\tau^2 (\xi_{i_p} - y)^2}\,\text{d}y & p = q.
\end{array}\right.
\end{align}
\begin{lem}\label{sinc_mumag}
After applying the substitution $\tau = \sinh(t)$ the transformed integral \eqref{matrix_op_sep} with integrand $\widetilde{f}(t) := \cosh(t) \, f(\sinh(t))$ allows an exponentially convergent 
sinc quadrature, cf. Theorem~\ref{sinc_theo}, where the constants in \eqref{quad_err}  depend on the parameters in \eqref{sep_integrands}.
\end{lem}
\begin{pf}
We assume w.l.o.g. $q=1$ and \eqref{sep_integrands} to be transformed to integrals over intervals $[a_p,\,b_p]$, i.e. $a_p := \xi_{i_p} - j_p h_p$ and $b_p := \xi_{i_p} - (j_p - 1)h_p$, where $h_p$ is the length of the interval $I_{j_p}^{(p)}$, which w.l.o.g. is assumed to be constant on the $p$-th axis. We set $C:= \prod_{j=1}^3 [a_j,b_j] \subset \prod_{j=1}^3 [h_j/2,\beta_j - \alpha_j]$ (assume midpoints as collocation points); then the transformed integrand in \eqref{matrix_op_sep} reads
\begin{align}\label{transformed_f}
\widetilde{f}(t) = \int_C  x^{(1)}\,\cosh(t)\,\sinh^2(t)\,\exp\big(-\sinh^2(t)\, \sum_{j=1}^3 {x^{(j)}}^2\big)\,\text{d}\big(x^{(1)},x^{(2)},x^{(3)}\big).
\end{align}
We obtain analytically up to constants
\begin{align}\label{f_anal}
\widetilde{f}(t) \propto \frac{\cosh(t)}{\sinh^2(t)}\Big(e^{-a_1^2\sinh^2(t)} - e^{-b_1^2\sinh^2(t)} \Big)\Big(\text{erf}\big(b_2 \sinh(t)\big) - \text{erf}\big(a_2\sinh(t)\big) \Big)\Big(\text{erf}\big(b_3 \sinh(t)\big) - \text{erf}\big(a_3\sinh(t)\big) \Big),
\end{align}
where the so-called error function is defined as
\begin{align}
\text{erf}(x) := \frac{2}{\sqrt{\pi}} \int_0^x e^{-\tau^2}\,\text{d}\tau.
\end{align}
The functions $\text{erf}\big(\sinh(z)\big)/\sinh(z)$, $\exp\big(-\sinh^2(z)\big)$ and $\cosh(z)$ are all entire functions, hence $\widetilde{f}$ is holomorphic over $\mathbb{C}$.\\
Moreover there holds ($a>0$)
\begin{align}
\exp\big(-a^2\sinh^2(t)\big) = \mathcal{O}\big(\exp(-\tfrac{a^2}{4} e^{2|t|})\big) \,\,\text{as}\,\, |t|\rightarrow \infty
\end{align} and using asymptotic expansion for the error function \footnote{$\text{erf}(x) \approx 1 - \frac{\exp(-x^2)}{\sqrt{\pi}x} \big( 1 - \frac{1}{2x^2} + \frac{3}{(2x^2)^2} - \frac{15}{(2x^2)^3} \pm \hdots\big),\, x \gg 1$} gives ($a,b>0,\, a\neq b$)
\begin{align}
\text{erf}\big(b \sinh(t)\big) - \text{erf}\big(a\sinh(t)\big) = \mathcal{O}\Big(\frac{\exp\big(-a^2\sinh^2(t)\big) - \exp\big(-b^2\sinh^2(t)\big)}{\sinh(t)} \Big) \,\,\text{as}\,\, |t|\rightarrow \infty,
\end{align} which shows the required double exponential decay for $\widetilde{f}$.\\
It remains to show that $N(\widetilde{f},D_\delta) < \infty$. We get for $z = t \pm i \delta$, $H:= \sum_{j=1}^3 h_j^2/4$ and $C_0 = \prod_{j=1}^3 [h_j/2,\beta_j - \alpha_j]$
\begin{align}
\int_{\mathbb{R}} \left|\int_C  x^{(1)}\, \cosh(t \pm i \delta)\,\sinh^2(t \pm i \delta)\,\exp\big(-\sinh^2(t \pm i \delta)\,\sum_{j=1}^3 {x^{(j)}}^2 \big)\,\text{d}\big(x^{(1)},x^{(2)},x^{(3)}\big) \right|\,\text{d}t & \,\leq \\[0.1cm]
\left|C_0\right|\,\int_{\mathbb{R}} \left|\cosh(t \pm i \delta)\right|\,\left|\sinh^2(t \pm i \delta)\right|\,\left|\exp\big(-\sinh^2(t \pm i \delta) H\big)\right|\,\text{d}t < \infty, & 
\end{align}
since the remaining integrand is a smooth function in $t$ with (double) exponential decay as $|t|\rightarrow \infty$.\\
This completes the proof.\hfill $\square$
%

\end{pf}
\begin{rmk}
Since in an estimate for $N(f,D_\delta)$ a term like $|\exp\big(-a^2\sinh^2(t \pm i\delta)\big)| = \exp\Big(-\tfrac{a^2}{2}\big(\cos(2\delta)\cosh(2t)-1\big)\Big)$ can have positive exponent (e.g. $t$ close to zero), the norm cannot be estimated uniformly in $h_p$, prohibiting an exponentially convergent sinc quadrature for $h_p \rightarrow 0$. Nevertheless, for the grid assumed to be fixed, Lemma~\ref{sinc_mumag} gives us a Kronecker product approximation of the operator \eqref{pot_op} with separation rank $R$, see Cor.~\ref{kron_op}, where $R = \mathcal{O}(|\log\epsilon|\cdot \log|\log\epsilon|)$ for a prescribed accuracy $\epsilon$. 
\end{rmk} 
\begin{cor}\label{kron_op}
The operator \eqref{pot_op} admits a Kronecker product approximation of the form \eqref{kronecker_form} with rank $R$, where $R = \mathcal{O}(|\log\epsilon|\cdot \log|\log\epsilon|)$ for a prescribed accuracy $\epsilon$.
\end{cor}
\begin{pf}
The substitution $\tau = \sinh(t)$ preserves the symmetry in the integrand \eqref{gauss_trans}, leading to a $R+1$-term sinc quadrature after applying Lemma~\ref{sinc_mumag}. Omitting the first term, which is zero, leads to the $R$-term representation for the potential operator $\mathcal{P}: \, \varprod_{q=1}^3 \bigotimes_{p=1}^3\mathbb{R}^{n_p} \rightarrow \bigotimes_{p=1}^3\mathbb{R}^{n_p}$ (cf. \eqref{pot_op} and \eqref{matrix_op_sep}) with 
\footnote{One notes that with $I = i_1 + (i_2 - 1)n_1 + (i_3 - 1)n_2^2$ and $J = j_1 + (j_2 - 1)n_1 + (j_3 - 1)n_2^2$ the entries of a matrix $A \in \mathbb{R}^{\prod_{p=1}^3 n_p \times \prod_{p=1}^3 n_p}$ given by $a_{IJ} = a_{i_1 j_1}^{(1)}a_{i_2 j_2}^{(2)}a_{i_3 j_3}^{(3)}$ correspond to the $(I,J)$-entry of $A^{(3)} \otimes A^{(2)} \otimes A^{(1)}$.} 
\begin{align}\label{kron_operator}
\mathcal{P}^{(q)} = -\frac{1}{2\pi^{\frac{3}{2}}}\sum_{k=1}^R \alpha_k\,U_k^{(q_3)} \otimes U_k^{(q_2)} \otimes U_k^{(q_1)},
\end{align}
with 
\begin{align}
\big(U_k^{(q_p)}\big)_{i_p j_p} =  h_{i_p j_p}^{(p)}\big(\sinh(t_k)\big)\quad \text{and} \quad \alpha_k = \cosh(t_k)\,\sinh^2(t_k),
\end{align}
where $t_k = k \vartheta$ for $\vartheta = c_0\,\log(R)/R$ and appropriate $c_0>0$ cf. Theorem~\ref{sinc_theo}.\hfill $\square$
\end{pf}
The evaluation of one component of $\mathcal{P}$ (cf. \eqref{kron_operator} for a rank$-1$ tensor, i.e. $\mathcal{X} = v^{(3)} \otimes v^{(2)} \otimes v^{(1)} \equiv \llbracket v^{(3)},v^{(2)},v^{(3)}\rrbracket\in \mathcal{C}_{\boldsymbol{n},1}$, is given as
\begin{align}\label{kron_eval}
\mathcal{P}^{(q)}\mathcal{X} = -\frac{1}{2\pi^{\frac{3}{2}}}\sum_{k=1}^R \alpha_k\,\big(U_k^{(q_3)}\,v^{(3)}\big) \otimes \big(U_k^{(q_2)}\,v^{(2)}\big) \otimes \big(U_k^{(q_1)}\,v^{(1)}\big),
\end{align} 
and amounts in a computational cost of $\mathcal{O}(R\sum_{p=1}^3 n_p^2)$.\\ 
It is not far to seek a reduction of this complexity by reducing the cost for the matrix-vector product (cf. Sec.~\ref{fft}) or reduction of the separation rank $R$ (cf. Sec.~\ref{recomp}).\\  
By using the relations\footnote{The symbol $\odot$ stands here for the \textit{Khatri-Rao product}, cf. Sec.~\ref{arithmetics}} \cite{bader_efficient_2008} (we assume appropriate dimensions for the involved matrices)
\begin{align}
\big(A_1 \otimes A_2 \big)\, \big(B_1 \odot B_2 \big) & \,= A_1B_1 \odot A_2B_2 \\[0.1cm]
\big(A_1 \otimes A_2 \big)\, \big(C_1 \otimes C_2 \big) & \,= A_1C_1 \otimes A_2C_2,
\end{align}
and 
\begin{align}
\text{vec}\Big( \llbracket \boldsymbol{\lambda};\,V^{(1)},V^{(2)},V^{(3)} \rrbracket \Big) = \big(V^{(3)} \odot V^{(2)} \odot V^{(1)}\big) \boldsymbol{\lambda}, 
\end{align}
respectively
\begin{align}
\text{vec}\Big( \llbracket \mathcal{C};\,V^{(1)},V^{(2)},V^{(3)} \rrbracket \Big) = \big(V^{(3)} \otimes V^{(2)} \otimes V^{(1)}\big) \text{vec}(\mathcal{C}), 
\end{align}
we get for the evaluation of \eqref{kron_operator} for $\mathcal{X} \in \mathcal{C}_{\boldsymbol{n},r}$ (also compare with \cite{exl_fast_2012_2})
\begin{align}\label{kron_eval_cp}
\mathcal{P}^{(q)}\mathcal{X} = -\frac{1}{2\pi^{\frac{3}{2}}}\sum_{k=1}^R \alpha_k\, \llbracket \boldsymbol{\lambda};\,U_k^{(q_1)}V^{(1)}, U_k^{(q_2)}V^{(2)},U_k^{(q_1)}V^{(3)} \rrbracket,
\end{align}
respectively for $\mathcal{X} \in \mathcal{T}_{\boldsymbol{n},\boldsymbol{r}}$ the formula
\begin{align}\label{kron_eval_tuck}
\mathcal{P}^{(q)}\mathcal{X} = -\frac{1}{2\pi^{\frac{3}{2}}}\sum_{k=1}^R \alpha_k\, \llbracket \mathcal{C};\,U_k^{(q_1)}V^{(1)}, U_k^{(q_2)}V^{(2)},U_k^{(q_1)}V^{(3)} \rrbracket.
\end{align}
\subsection{Some practical issues}\label{recomp}
\begin{figure}[hbtp]
\center
\includegraphics[scale = 0.4]{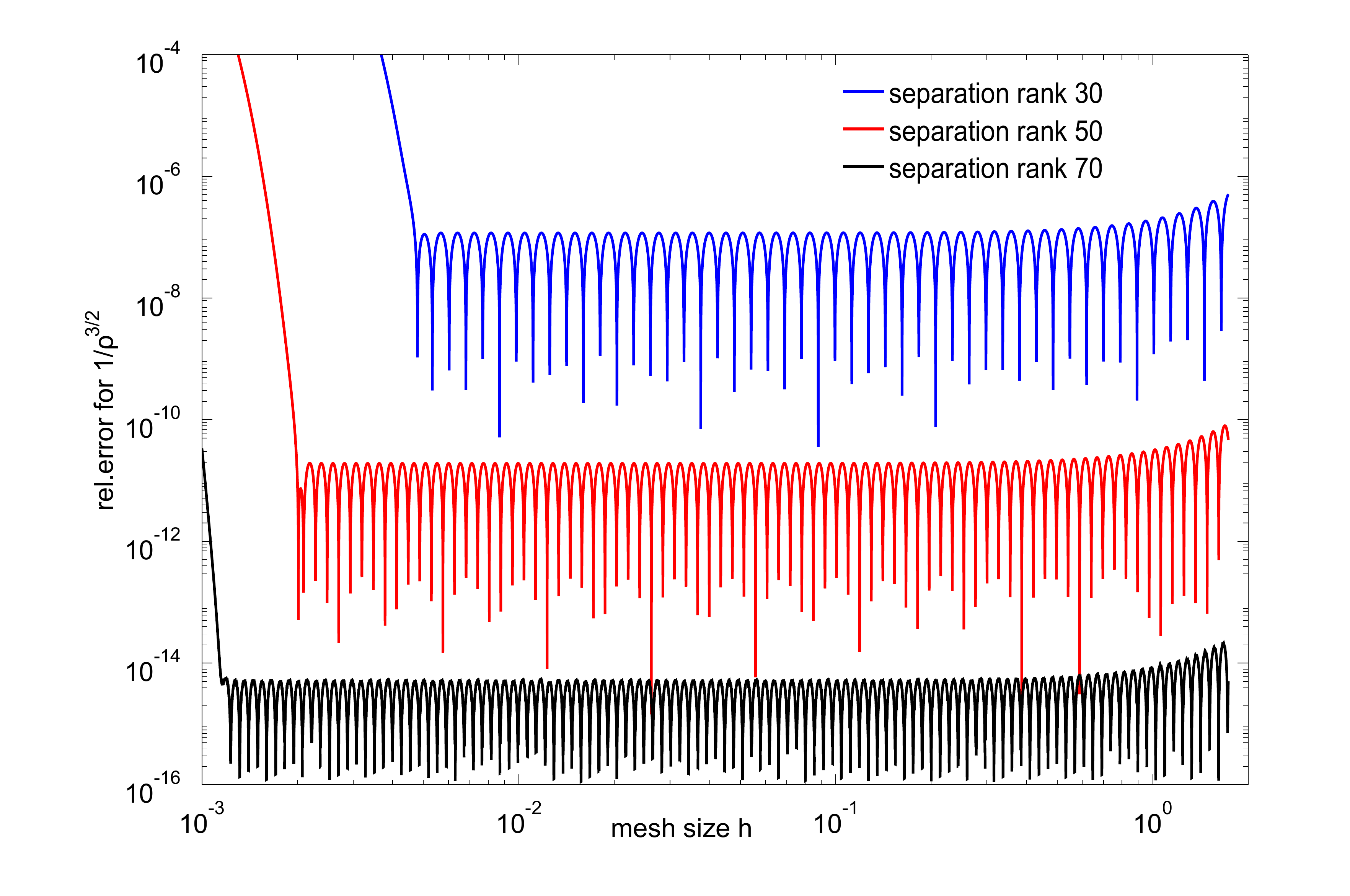}    
\caption{Relative error for sinc quadrature of \eqref{gauss_trans} with substitutuion $\tau = \sinh(t)$. $c_0 = 2.15$}\label{fig1}
\end{figure}

Here we briefly address the question how to choose the rank $R$.\\
If we apply sinc quadrature to the Gaussian transformed kernel \eqref{gauss_trans}, we can adaptively determine the rank by contolling the relative error of the quadrature in an interval $\rho \in [\rho_{\text{min}},\rho_{\text{max}}]$ corresponding to the mesh size parameter $h$ by the relation $\rho = \left\|\boldsymbol{\xi}_{\boldsymbol{i}} - \boldsymbol{y}\right\|^2 \geq 3h^2/4$ (cf. proof of Lemma~\ref{sinc_mumag} Eq. \eqref{transformed_f}). Since we can scale the computational domain to unity, we considered in Fig.~\ref{fig1} $\rho_{\text{max}} \leq 3$, corresponding to $\Omega_{\text{scaled}} = [0,1]^3$. We observe an uniform relative error bound for $h$ greater some $h_{\text{min}}$. Also compare with \cite{exl_fast_2012_2}. Moreover, we see from Fig.~\ref{fig1} the exponential decay of the error w.r.t. the rank (the logarithmic error is a decreasing affine function of the rank).\\
%
 The separation rank can also be reduced by applying recompression of the CP representation (cf. Sec.~\ref{Canonical}) corresponding to \eqref{kron_operator}, either by compressing \eqref{kron_operator} as Tucker tensor by Alg.~\ref{HOOI} and subsequent approximation of the resulting core to CP by optimization based algorithms \cite{acar_scalable_2011} yielding a canonical tensor with smaller rank, or direct CP approximation to a smaller rank. The resulting CP representation again corresponds to a Kronecker product representation. 
\section{FFT acceleration}\label{fft}
Here we address the question how to use FFT in order to reduce compuational costs for evaluating the operator $\mathcal{P}$.
\subsection{Multidimensional DFT for structured tensors}\label{DFT}
Again we present the following for the case of order-$3$ tensors; everthing is also valid for the general case of order-$d$.\\ 
For a given  tensor $\mathcal{X} \in \bigotimes_{p=1}^3\mathbb{R}^{n_p}$ the 3-d \textit{discrete Fourier Transform} (DFT) results in the complex tensor $\widehat{\mathcal{X}} \in \bigotimes_{p=1}^3\mathbb{C}^{n_p}$ and is defined as
\begin{align}\label{dft}
\widehat{x}_{k_1 k_2 k_3} = \sum_{j_1 = 1}^{n_1} \omega_{n_1}^{(k_1-1)(j_1-1)} \sum_{j_2 = 1}^{n_2} \omega_{n_2}^{(k_2-1) (j_2-1)} \sum_{j_3 = 1}^{n_3} \omega_{n_3}^{(k_3-1) (j_3-1)} x_{j_1 j_2 j_3},
\end{align}
where $\omega_{n_p}^{k_p j_p} = e^{-2\pi i \frac{k_p j_p}{n_p}}$.\\
The inverse discrete Fourier Transformation (IDFT) is given by
\begin{align}\label{idft}
x_{j_1 j_2 j_3} = \frac{1}{n_1 n_2 n_3}\sum_{k_1 = 1}^{n_1} \omega_{n_1}^{-(k_1-1)(j_1-1)} \sum_{k_2 = 1}^{n_2} \omega_{n_2}^{-(k_2-1) (j_2-1)} \sum_{k_3 = 1}^{n_3} \omega_{n_3}^{-(k_3-1) (j_3-1)} \widehat{x}_{k_1 k_2 k_3}.
\end{align}
FFT variants with zero-padding are commonly used because of their almost linear scaling, i.e. $\mathcal{O}(\sum_{p=1}^3 \log(n_p)\prod_{q=1}^3 n_q )$.\\
The convolution theorem holds, i.e. in multi-index notation and using the operator symbols $\mathcal{F}$ and $\mathcal{F}^{-1}$ for the DFT and IDFT respectively
\begin{align}
\mathcal{X} \ast \mathcal{K}_{\mathbf{n}} = \mathcal{F}^{-1} \big(\mathcal{F}(\mathcal{X})\cdot \mathcal{F}(\mathcal{K})\big),
\end{align}
where the subscript denotes the $\mathbf{n}$- shift and $\cdot$ the elementwise product (\textit{Hadamard product}).\\
The following is straight forward to show.
\begin{lem}\label{DFT_cp}
For a canonical tensor $\mathcal{X} = \llbracket\, \boldsymbol{\lambda};\, U^{(1)},U^{(2)},U^{(3)} \,\rrbracket \in \mathcal{C}_{\boldsymbol{n},r},\,\, x_{j_1 j_2 j_3} = \sum_{l=1}^r \lambda_j \,u_{j_1 l}^{(1)}\, u_{j_2 l}^{(2)}\, u_{j_3 l}^{(3)}$ the DFT is given by
\begin{align}\label{dft_cp}
\widehat{\mathcal{X}} = \llbracket\, \boldsymbol{\lambda};\, \mathcal{F}_{\text{1d}}\big({U}^{(1)}\big),\mathcal{F}_{\text{1d}}\big({U}^{(2)}\big),\mathcal{F}_{\text{1d}}\big({U}^{(3)}\big) \,\rrbracket,
\end{align} where the (1-d) Fourier transform $\mathcal{F}_{\text{1d}}$ is only taken along each column of a factor matrix.\\
The IDFT is given as
\begin{align}\label{idft_cp}
\mathcal{X} = \llbracket\, \boldsymbol{\lambda};\, \mathcal{F}_{\text{1d}}^{-1}\big(U^{(1)}\big),\mathcal{F}_{\text{1d}}^{-1}\big(U^{(2)}\big),\mathcal{F}_{\text{1d}}^{-1}\big(U^{(3)}\big) \,\rrbracket.
\end{align} 
\end{lem}
\begin{pf}
\begin{align}
\widehat{x}_{k_1 k_2 k_3} = \sum_{l=1}^r \lambda_r \Big(\sum_{j_1 = 1}^{n_1} \omega_{n_1}^{(k_1-1) (j_1-1)}\, u_{j_1 l}^{(1)}\Big) \Big(\sum_{j_2 = 1}^{n_2} \omega_{n_2}^{(k_2-1) (j_2-1)}\, u_{j_2 l}^{(2)} \Big) \Big(\sum_{j_3 = 1}^{n_3} \omega_{n_3}^{(k_3-1) (j_3-1)}\, u_{j_3 l}^{(3)}\Big)
\end{align} and analog for the IDFT.\hfill $\square$
\end{pf}
\begin{lem}\label{DFT_tucker}
For a Tucker tensor $\mathcal{X} = \llbracket\, \mathcal{C};\, U^{(1)},U^{(2)},U^{(3)} \,\rrbracket \in \mathcal{T}_{\boldsymbol{n},\boldsymbol{r}}, \,\, x_{j_1 j_2 j_3} = \sum_{j_1^\prime,j_2^\prime,j_3^\prime=1}^{r_1,r_2,r_3} c_{j_1^\prime j_2^\prime j_3^\prime} \,u_{j_1 j_1^\prime}^{(1)}\, u_{j_2 j_2^\prime}^{(2)}\, u_{j_3 j_3^\prime}^{(3)}$ the DFT is given by
\begin{align}\label{dft_t}
\widehat{\mathcal{X}} = \llbracket\, \mathcal{C};\, \mathcal{F}_{\text{1d}}\big(U^{(1)}\big),\mathcal{F}_{\text{1d}}\big(U^{(2)}\big),\mathcal{F}_{\text{1d}}\big(U^{(3)}\big) \,\rrbracket,
\end{align} where the (1-d) Fourier transform $\mathcal{F}_{\text{1d}}$ is only taken along each column of a factor matrix.\\
The IDFT is given as
\begin{align}\label{idft_t}
\mathcal{X} = \llbracket\, \mathcal{C};\, \mathcal{F}_{\text{1d}}^{-1}\big(U^{(1)}\big),\mathcal{F}_{\text{1d}}^{-1}\big(U^{(2)}\big),\mathcal{F}_{\text{1d}}^{-1}\big(U^{(3)}\big) \,\rrbracket.
\end{align} 
\end{lem}
\begin{pf}
\begin{align}
\widehat{x}_{k_1 k_2 k_3} = \sum_{j_1^\prime,j_2^\prime,j_3^\prime=1}^{r_1,r_2,r_3}  c_{j_1^\prime j_2^\prime j_3^\prime} \Big(\sum_{j_1 = 1}^{n_1} \omega_{n_1}^{(k_1-1) (j_1-1)}\, u_{j_1 j_1^\prime}^{(1)}\Big) \Big(\sum_{j_2 = 1}^{n_2} \omega_{n_2}^{(k_2-1) (j_2-1)}\, u_{j_2 j_2^\prime}^{(2)} \Big) \Big(\sum_{j_3 = 1}^{n_3} \omega_{n_3}^{(k_3-1) (j_3-1)}\, u_{j_3 j_3^\prime}^{(3)}\Big)
\end{align} and analog for the IDFT.\hfill $\square$
\end{pf}
The DFT as well as the IDFT for structured tensors is therefor basically one-dimensional, more precise, assuming FFT and IFFT implementations, the costs are $\mathcal{O}(R\sum_p n_p \log n_p)$ for canonical tensors and $\mathcal{O}(\sum_p r_p n_p \log n_p)$ for Tucker tensors respectively. This matches asymptotically the costs for one dimensional DFFT times the rank constants.
\subsection{Convolution kernel and FFT evaluation}
The evaluation of the operator \eqref{kron_operator} scales quadratically in the number of collocation points in one dimension. Even though this is super-optimal, also refered to as \textit{sub-linear} (cf. \cite{exl_fast_2012_2} or (slightly misleading) as \textit{super-linear} (cf. \cite{goncharov_2010}), we can still reduce this complexity (on uniform grids) by observing a convolution in \eqref{intop2}.\\
Let us assume that $h_p$ is constant for each $p$ and the collocation points to be the midpoints of the intervals $I_{j_p}^{(p)} \equiv I^{(p)}$.\\
By applying the substitution used in the proof of Lemma~\ref{sinc_mumag} we get for the functions in \eqref{sep_integrands}
\begin{align}
h_{i_p - j_p}^{(p)}(\tau) = h_{i_p j_p}^{(p)}(\tau) = 
\left\{\begin{array}{l l}
\int_{a_{i_p-j_p}}^{b_{i_p-j_p}} e^{-\tau^2 x^2}\,\text{d}x & p \neq q, \\*[\jot]
\int_{a_{i_p-j_p}}^{b_{i_p-j_p}} x\,e^{-\tau^2 x^2}\,\text{d}x & p = q,
\end{array}\right.
\end{align}  
where $a_{i_p-j_p} = (i_p - j_p)h_p - \tfrac{h_p}{2}$ and $b_{i_p-j_p} = (i_p - j_p)h_p + \tfrac{h_p}{2}$.\\ 
For $i_p,j_p = 1 \hdots n_p$ these are $2n_p - 1$ different integrals. We therefor set $i_p - j_p = J_p = 1\hdots 2n_p - 1$ and identify the \textit{convolution kernel} $\boldsymbol{G}_{\boldsymbol{i}-\boldsymbol{j}}^{(q)} \in \bigotimes_{p=1}^3 \mathbb{R}^{{2n_p - 1}}$ with entries 
\begin{align}\label{conv_op}
G_{J_1 J_2 J_3}^{(q)} = \frac{2}{\sqrt{\pi}} \int_{\mathbb{R}} \tau^2 \prod_{p=1}^3 h_{J_p}^{(p)}(\tau) \,\text{d}\tau.
\end{align} 
Lemma~\ref{sinc_mumag} also holds for \eqref{conv_op} (after the substitution $\tau = \sinh(t)$) and hence we can apply sinc quadrature leading to (cf. \eqref{pot_op} and \eqref{matrix_op_sep}) the \textit{canonical representation}
\begin{align}\label{conv_pot}
\mathcal{P}^{(q)} = -\frac{1}{2\pi^{\frac{3}{2}}}\llbracket\,\boldsymbol{\lambda};\, D_1^{(q)},D_2^{(q)},D_3^{(q)} \,\rrbracket \in \mathcal{C}_{2\boldsymbol{n}-1,R},
\end{align}
with 
\begin{align}
\big(D_p^{(q)}\big)_{J_p\, k} =  h_{J_p}^{(p)}\big(\sinh(t_k)\big)\quad \text{and} \quad \lambda_k = \cosh(t_k)\,\sinh^2(t_k),
\end{align}
where $t_k = k \vartheta$ for $\vartheta = c_0\,\log(R)/R$ and appropriate $c_0>0$ cf. Theorem~\ref{sinc_theo}.\\
$\mathcal{P}^{(q)}$ is a \textit{separable convolution kernel}.\\
By denoting the elementwise product (\textit{Hadamard product}) for tensors with $\cdot$ and using DFT, the evaluation of one component of $\mathcal{P}$ (cf. \eqref{conv_pot}) for a tensor $\mathcal{X} \in \bigotimes_{p=1}^3 \mathbb{R}^{n_p}$ with (zero-padded) Fourier transform $\widehat{\mathcal{X}} \in \bigotimes_{p=1}^3 \mathbb{C}^{{2n_p - 1}}$ reads
\begin{align}\label{fft_pot}
\mathcal{P}^{(q)}\ast\mathcal{X} = \mathcal{F}^{-1} \Big(-\frac{1}{2\pi^{\frac{3}{2}}}\llbracket\,\boldsymbol{\lambda};\, \widehat{D}_1^{(q)},\widehat{D}_2^{(q)},\widehat{D}_3^{(q)} \,\rrbracket \cdot \widehat{\mathcal{X}}\Big)
\end{align}
The complexity of FFT-based convolution in \eqref{fft_pot} with an unstructured tensor $\mathcal{X}$ is dominated by the costs for the FFT of $\mathcal{X}$, and amounts in a complexity of $\mathcal{O}(R\sum_{p=1}^3 \log n_p \prod_{q=1}^3 n_q)$, which is comparable with usual FFT-based computation of the scalar potential (except the constant $R$, the order of sinc quadrature), cf. \cite{abert_2013},\cite{abert_2012}.\\
For structured tensors $\mathcal{X}$ the complexity of FFT-based convolution with the separable kernels \eqref{conv_pot} scales like $n_p\log(n_p)$ in the mode sizes due to the Lemmas \ref{DFT_cp} and \ref{DFT_tucker} of Sec.\ref{DFT}.\\ 
We can think of recompression of \eqref{conv_pot}, e.g. efficient compression of a CP tensor to a Tucker tensor by Alg.\ref{HOOI} or to the \textit{Tensor Train format} (TT) \ref{TTtensor}, if the magnetization itself is represented in Tucker representation. The (complex) Hadamard product in Fourier space can be performed efficiently with the techniques described in Sec.~\ref{Hadamard_cp} or \ref{Hadamard_tucker}, depending on the structure of the magnetization component tensors.\\
Choosing a large number of quadrature terms ($\propto$ rank $R$) becomes a feasible option, which results in an accurate representation of the operator \eqref{pot_op}. Since the computation of \eqref{pot_op} as well as the recompression of the discrete kernels \eqref{conv_pot} are done in a setup-phase in a micromagnetic simulation, the higher amount of work due to higher ranks does not significantly influence the overall computing time. \\
%
%
\begin{algorithm}
\caption{Scalar potential DFFT; \texttt{potential($\mathcal{M}^{(1)},\mathcal{M}^{(2)},\mathcal{M}^{(3)},\,\boldsymbol{h},\,R,\,c_0 ,\, \texttt{tol} > 0$)}}
\label{scpot_fft_tucker}
\begin{algorithmic}
\Require $\mathcal{M}^{(p)}\in\mathcal{T}_{\mathbf{n},\mathbf{r}^{(p)}}, \, h_p > 0\, (p = 1\hdots 3),\, R\in\mathbb{N},\,c_0 > 0, \texttt{tol} > 0$
\Ensure $\Phi\in \mathcal{T}_{\mathbf{n},\mathbf{r}^\prime}$\\
\textbf{Setup}
\begin{itemize}
	\item \text{Compute the CP kernels \eqref{conv_pot}}
  \item \text{Compress the kernels to Tucker tensors with tolerance \texttt{tol}}
  \item \text{Compute DFFT of Tucker kernels by using Lemma \ref{DFT_tucker}}
\end{itemize}
\textbf{Actual computation}
\For{$q = 1 \hdots 3$}
\begin{itemize}
\item \text{Compute DFFT of $q$-th Tucker magnetization component by using Lemma \ref{DFT_tucker}}
\item \text{Compute Tucker Hadamard products of magnetization component and kernels in Fourier space} \text{using e.g. the tolerance \texttt{tol}}
\item \text{Compute IDFFT for result of previous step}
\end{itemize}
\EndFor
\State \text{Perform approximate summation of results in previous step using e.g. the tolerance \texttt{tol}}
\end{algorithmic}
\end{algorithm} 

\begin{figure}[hbtp]
\includegraphics[scale = 0.50]{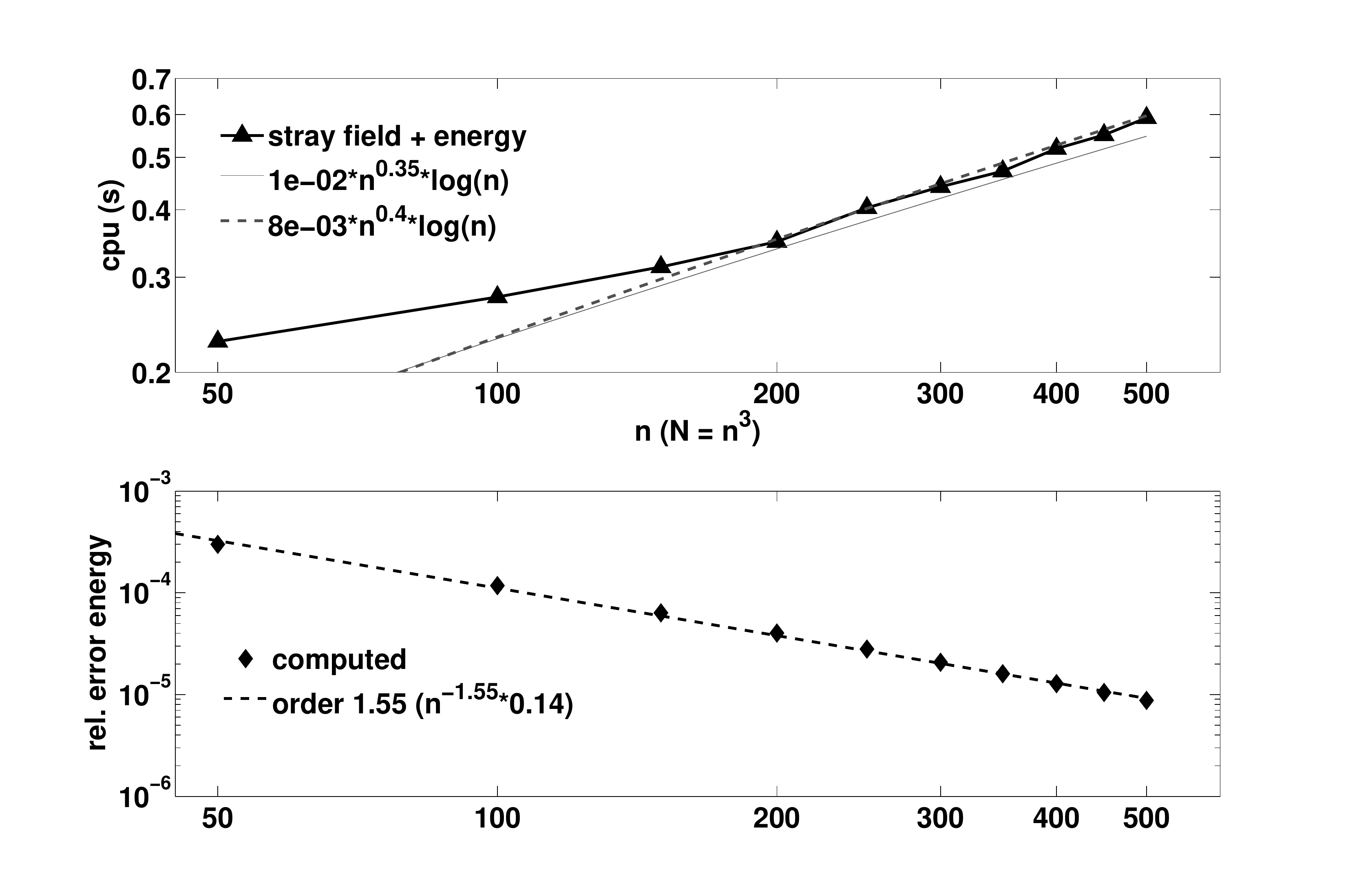}    
\caption{Complexity and relative error in the energy for Alg.~\ref{scpot_fft_tucker}.}\label{fig2}
\end{figure}
\begin{figure}[hbtp]
\center
\includegraphics[scale = 0.50]{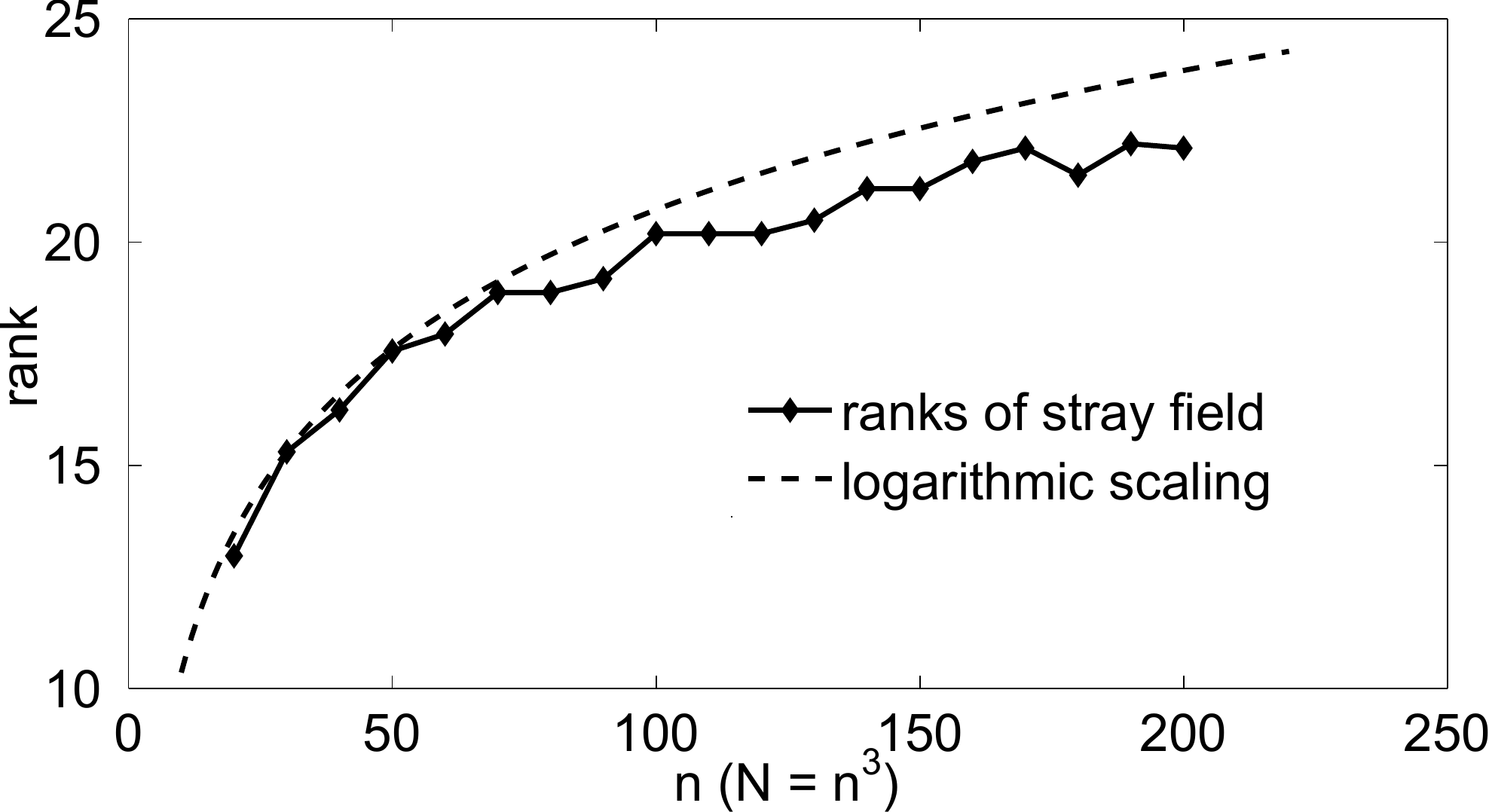}    
\caption{Ranks of the computed stray field using Alg.~\ref{scpot_fft_tucker} with accuracy $1$e-$12$ and a flower-like magnetization state.}\label{fig3}
\end{figure}
Alg.~\ref{scpot_fft_tucker} describes the FFT-based procedure for computing the scalar potential for Tucker magnetization.
Fig.~\ref{fig2} shows results on complexity and accuracy using Alg.~\ref{scpot_fft_tucker} for the case of uniform magnetization, cf. \cite{exl_fast_2012_2}. We observe the quasi linear complexity ($n\log n$), while the relative error in the energy decreases with order about $1.5$.\footnote{The discretization for the energy as well as the stray field calculation from the potential were carried out second order.} Fig.~\ref{fig3} shows logarithmic rank-grows for the stray field (averaged over modes) induced by a non-trivial (flower-like) magnteization state \cite{abert_2013} while using an accuracy of $1$e-$12$ in Alg.~\ref{scpot_fft_tucker}. Computations were performed in parallel in Matlab $7.11.0$ using the classes provided by the \textit{Tensor Toolbox} V.$2.5$ \cite{TTB_Software} on a Linux Workstation with a Quad-Core Intel i7 processor and 6 GB RAM.
\section{Conclusion} 
We gave a detailed review of tensor formats and their use in approximation of analytical operators.
A quadratically convergent collocation scheme on tensor product grids for the micromagnetic potential operator is proved to have a Kronecker product approximation with exponential convergence in the separation rank. This is a mathematically rigorous confirmation of the algorithm given in \cite{exl_fast_2012_2}. The discrete Fourier transform for structured tensors can be used for the purpose of accelerating the method on uniform grids, yielding quasi linear complexity in the number of collocation points in one dimension. 
\appendix
\section{Best approximation}\label{sec_best_appr}
\begin{defi}
Let $(V,d)$ be a metric space and $\emptyset \neq U \subseteq V$. A best approximation of $v \in V$ in $U$ is an element $u^\ast \in U$ such that
\begin{align}\label{best_appr}
d(u^\ast,v) = d(U,v) := \inf\{d(u,v):\,u\in U\}, 
\end{align}
i.e. the infimum is attained in $U$.
\end{defi}
If $V$ is a normed vector space the condition \eqref{best_appr} reads: $\left\|\,v - u^\ast\,\right\| \leq \left\|\,v - u\,\right\|$ for all $u \in U$.
\begin{lem}\label{best_approx_lemma}
Let $(V,\left\|\,.\,\right\|)$ be a normed vector space with $\text{dim}(V)<\infty$ and $\emptyset \neq U \subseteq V$ a closed subset. Then, for all $v\in V$ there exists an element $u_{\text{best}}\in U$ with $\left\|\,v - u_{\text{best}}\,\right\| \leq \left\|\,v - u\,\right\|$ for all $u \in U$.
\end{lem}
\begin{pf}
Let $v\in V$ and $\widetilde{u}\in U$ and define $D:= U \cap \{u \in V:\, \left\| v - u\,\right\| \leq \left\|\,v - \widetilde{u}\,\right\|\} \subseteq U$. 
Since $D$ is non-empty ($\widetilde{u} \in D$), bounded (triangle inequality) and closed ($D$ is intersection of two closed sets) and hence compact\footnote{This conclusion fails in infinite dimensional normed vector spaces in general.}, the continuity of the function $f:D \rightarrow \mathbb{R},\, f(u) := \left\|\,v - u\,\right\|$ together with the extrem value theorem\footnote{The image of compact sets under continuous functions is compact.} ensures that $f$ attains its minimal value at $u_{\text{best}}\in D \subseteq U$. \hfill $\square$
\end{pf}
\section{Hadamard product for structured tensors}\label{arithmetics}
\subsection{CP tensors}\label{Hadamard_cp}
For two canonical tensors $\mathcal{X} = \llbracket\, \boldsymbol{\lambda};\, U^{(1)},U^{(2)},U^{(3)} \,\rrbracket \in \mathcal{C}_{\boldsymbol{n},r}$ and $\mathcal{Y} = \llbracket\, \boldsymbol{\mu};\, V^{(1)},V^{(2)},V^{(3)} \,\rrbracket \in \mathcal{C}_{\boldsymbol{n},r^\prime}$ of equal mode sizes the \textit{Hadamard product} (elementwise product) is
\begin{align}\label{cp_had}
\big(\mathcal{X}\cdot\mathcal{Y}\big)_{IJK} = \sum_{i}\sum_{j} \lambda_{i}\,\mu_{j}\,\big( u_{Ii}^{(1)}v_{Ij}^{(1)}\big)\,\big( u_{Ji}^{(2)}v_{Jj}^{(2)}\big)\,\big( u_{Ki}^{(3)}v_{Kj}^{(3)}\big) \equiv 
\llbracket\, \boldsymbol{\nu};\, W^{(1)},W^{(2)},W^{(3)}\,\rrbracket \in \mathcal{C}_{\boldsymbol{n},rr^\prime}, 
\end{align}
where $\boldsymbol{\nu} = [\boldsymbol{\lambda},\boldsymbol{\mu}]$ and $W^{(p)} = \left[u_1^{(p)} \cdot v_1^{(p)}\, | \hdots | u_1^{(p)} \cdot v_{r^\prime}^{(p)}\,| \hdots u_r^{(p)} \cdot v_{r^\prime}^{(p)}\right]$. The cost for forming \eqref{cp_had} is of order $\mathcal{O}(r r^\prime\sum_{p=1}^3 n_p )$.
The new CP tensor has rank $r \,r^\prime$. Direct recompression can be considered e.g. by optimization \cite{acar_scalable_2011}.
\subsection{Tucker tensors}\label{Hadamard_tucker}
For two Tucker tensors $\mathcal{X} = \llbracket\, \mathcal{C};\, U^{(1)},U^{(2)},U^{(3)} \,\rrbracket \in \mathcal{T}_{\boldsymbol{n},\boldsymbol{r}}$ and $\mathcal{Y} = \llbracket\, \mathcal{D};\, V^{(1)},V^{(2)},V^{(3)} \,\rrbracket \in \mathcal{T}_{\boldsymbol{n},\boldsymbol{r}^\prime}$ of equal mode sizes the \textit{Hadamard product} (elementwise product) is
\begin{align}\label{t_had}
\big(\mathcal{X}\cdot\mathcal{Y}\big)_{IJK} = \sum_{i,j,k}\sum_{l,m,n} c_{ijk}\,d_{lmn}\,\big( u_{Ii}^{(1)}v_{Il}^{(1)}\big)\,\big( u_{Jj}^{(2)}v_{Jm}^{(2)}\big)\,\big( u_{Kk}^{(3)}v_{Kn}^{(3)}\big). 
\end{align}
The Hadamard product \eqref{t_had} can be written in compact form with the \textit{Khatri-Rao product}.
\begin{defi}
Given two matrices $A\in \mathbb{R}^{I \times K}$ and $B\in \mathbb{R}^{J \times K}$, their Khatri-Rao product is given by
\begin{align}
A \odot B = \left[ a_{:1} \otimes b_{:1} \quad a_{:2} \otimes b_{:2} \quad \hdots \quad a_{:K} \otimes b_{:K} \right] \in \mathbb{R}^{IJ \times K}.
\end{align}
\end{defi} 
It is straight forward to show
\begin{align}\label{t_had2}
\mathcal{X}\cdot\mathcal{Y} =  \llbracket\, \mathcal{E};\,{\big({V^{(1)}}^T \odot {U^{(1)}}^T\big)}^T,{\big({V^{(2)}}^T \odot {U^{(2)}}^T\big)}^T,{\big({V^{(3)}}^T \odot{U^{(3)}}^T\big)}^T \,\rrbracket \in \mathcal{T}_{\boldsymbol{n},\boldsymbol{r}\cdot \boldsymbol{r}^\prime},
\end{align} where $\mathcal{E}$ is the reshaped tensor product of $\mathcal{C}$ and $\mathcal{D}$, i.e. $e_{(il)(jm)(km)} = c_{ijk}\,d_{lmn}$. The costs for computing \eqref{t_had2} are therefor $\mathcal{O}(\sum_{p=1}^3 n_pr_p r_p^\prime + \prod_{p=1}^3 r_p r_p^\prime)$.\\
The new Tucker tensor has ranks $\boldsymbol{r}\cdot \boldsymbol{r}^\prime$. It is therefore practical to recompress the original cores before building the tensor-product; also recompression of \eqref{t_had2} is highly recommended if further operations with $\mathcal{X}\cdot\mathcal{Y}$ are planned. 
Indeed, in practice often very effective recompression of the core $\mathcal{E}$ and/or Hadamard product itself is observed. 
%
\section{Relation between Tucker tensors and Tensor Trains (TT) in 3 dimensions}\label{TTtensor}
A Tensor Train $\mathcal{A}$ (TT) \cite{oseledets_2011} in $d$ dimensions is given as
\begin{align}\label{tt1}
a_{i_1 i_2 \hdots i_d} = G_1(i_1)G_2(i_2)\hdots G_d(i_d),
\end{align}
where $G_k(i_k)$ is an $r_{k-1}\times r_{k}$ matrix with $r_0 = r_d = 1$. Writing out the products leads to
\begin{align}\label{tt2}
a_{i_1 i_2 \hdots i_d} = \sum_{\alpha_1, \hdots, \alpha_{d-1}} G_1(i_1, \alpha_1)G_2(\alpha_1,i_2,\alpha_2)\hdots G_{d-1}(\alpha_{d-2} i_{d-1} \alpha_{d-1} ) G_d(\alpha_{d-1}, i_d).
\end{align}
There holds a quasi-best approximation result due to best rank-$r_k$ approximation of the unfolding matrices of $\mathcal{A}$ \cite{oseledets_2011}. Recompression (\textit{rounding}) and CP2TT is also possible \cite{oseledets_2011}, 
as well as black-box approximation by ACA type algorithms \cite{tt_tensor2}.\\
In three dimensions \eqref{tt2} reduces to
\begin{align}\label{tt_3d_1}
a_{i_1 i_2 i_3} = \sum_{\alpha_1,\alpha_2} G_1(i_1, \alpha_1)G_2(\alpha_1,i_2,\alpha_2)G_3(\alpha_{2}, i_3).
\end{align}
From \eqref{tt_3d_1} we conclude
\begin{align}\label{tt_3d_2}
\mathcal{A} = \mathcal{G} \times_1 G_1 \times_3 G_3^T \equiv \llbracket\, \mathcal{G}; G_1, \text{id}, G_{3}^T \,\rrbracket, 
\end{align}
where $\mathcal{G}$ denotes the $3$-tensor $G_2 \in \bigotimes_{p=1}^3\mathbb{R}^{m_p}$, where $m_p = r_p,\, p\neq2$ and $m_2 = n_2$. Recompression of \eqref{tt_3d_2} leads to a $(r_1^\prime,r_2^\prime,r_3^\prime)$-Tucker tensor.\\
The representation \eqref{tt_3d_2} is also known as \textit{Tucker2} decomposition of a tensor \cite{kolda_tensor_2009}.\\
On the other hand a Tucker tensor can also be easily converted to a TT, i.e. by mode-multiplication of one factor matrix (e.g. the second factor) we immediately get the form \eqref{tt_3d_2}. Recompression by TT-rounding can be done afterwards.
The Hadamard product (as well as many other arithmetic operations) can be carried out efficiently in TT-format, \cite{oseledets_2011}.
\section*{Acknowledgment}
Financial support by the Austrian Science Fund (FWF) SFB ViCoM (F4112-N13) and 
the Deutsche Forschungsgemeinschaft via the Graduiertenkolleg 1286 ``Functional Metal-Semiconductor Hybrid Systems''
is gratefully acknowledged.\\
The first author wants to thank Prof. W. Hackbusch for a helpful discussion on this topic.
\bibliographystyle{plainnat}
\bibliography{bibref}

\end{document}